\documentclass{article}

\usepackage{xcolor}
\usepackage{makecell}
\usepackage{multirow}
\usepackage{PRIMEarxiv}
\usepackage{natbib}
\usepackage[utf8]{inputenc} 
\usepackage[T1]{fontenc}    
\usepackage{hyperref}       
\usepackage{url}            
\usepackage{booktabs} 
\usepackage[ruled,linesnumbered]{algorithm2e}
\usepackage{amsmath}
\usepackage{amsfonts}       
\usepackage{nicefrac}       
\usepackage{microtype}
\usepackage{lipsum}
\usepackage{fancyhdr}       
\usepackage{graphicx}       
\graphicspath{{media/}}     
\newcommand{\bel}{\begin{eqnarray}\label}
\newcommand{\eel}{\end{eqnarray}}
\newcommand{\bes}{\begin{eqnarray*}}
\newcommand{\ees}{\end{eqnarray*}}
\newcommand{\bei}{\begin{itemize}}
\newcommand{\beiftnt}{\begin{itemize}\footnotesize}
\newcommand{\eei}{\end{itemize}}

\def\benu{\begin{enumerate}}
\def\eenu{\end{enumerate}}

\def\argmin{\mathop{\rm arg\, min}}

\def\complex{\mathop{{\rm I}\kern-.58em\hbox{\rm C}}\nolimits}

\def\Cov{\hbox{\rm Cov}}

\def\Var{\hbox{\rm Var}}

\def\mathbold{\boldsymbol} 

\def\ba{\mathbold{a}}

\def\bA{\mathbold{A}}

\def\bb{\mathbold{b}}

\def\hbC{{\widehat{\bfC}}}

\def\bC{\mathbold{C}}
\def\hbC{{\widehat{\bC}}}

\def\bI{\mathbold{I}}

\def\bx{\mathbold{x}}

\def\bX{\mathbold{X}}\def\tbX{{\widetilde{\bX}}}

\def\bz{\mathbold{z}}

\def\bZ{\mathbold{Z}}

\def\balpha{\mathbold{\alpha}}\def\halpha{\widehat{\alpha}}

\def\hbalpha{{\widehat{\balpha}}}\def\tbalpha{{\widetilde{\balpha}}}

\def\bbeta{\mathbold{\beta}}\def\hbeta{\widehat{\beta}}

\def\hbbeta{{\widehat{\bbeta}}}\def\tbbeta{{\widetilde{\bbeta}}}

\def\bGamma{\mathbold{\Gamma}}

\def\btheta{\mathbold{\theta}}

\def\hbtheta{{\widehat{\btheta}}}\def\tbtheta{{\widetilde{\btheta}}}

\def\lam{\lambda}

\def\bLambda{\mathbold{\Lambda}}

\def\bSigma{\mathbold{\Sigma}}\def\hbSigma{{\widehat{\bSigma}}}

\def\bmB{\mathbf{\mathcal{B}}}
\def\bmA{\mathbf{\mathcal{A}}}
\def\bmC{\mathbf{\mathcal{C}}}

\def\bmX{\mathbf{\mathcal{X}}}

\def\argmin{\operatorname{argmin} \displaylimits}

\def\mR{\mathcal{R}}

\def\T{\top}

\def\tvec{\text{vec}}

\def\bmB{\mathbf{\mathcal{B}}}
\def\bmA{\mathbf{\mathcal{A}}}
\def\bmC{\mathbf{\mathcal{C}}}

\def\bmX{\mathbf{\mathcal{X}}}

\def\argmin{\operatorname{argmin} \displaylimits}

\def\mR{\mathcal{R}}

\def\T{\top}

\def\tvec{\text{vec}}

\title{A General Framework of Brain Region Detection and Genetic Variants Selection in Imaging Genetics
}

\author{
  Siqiang Su \\
  Department of Statistics \& Actuarial Science \\
  The University of Hong Kong \\
  \texttt{u3008662@connect.hku.hk} \\
   \And
  Zhenghao Li \\
  Department of Statistics \& Actuarial Science  \\
  The University of Hong Kong \\
  \texttt{lizh@connect.hku.hk} \\
   \AND
   Long Feng \\
   Department of Statistics \& Actuarial Science \\
   The University of Hong Kong \\
   \texttt{lfeng@hku.hk} \\
   \And
   Ting Li \\
   Department of Applied Mathematics \\
   The Hong Kong Polytechnic University \\
   \texttt{tingeric.li@polyu.edu.hk} \\
}

\begin{document}
\maketitle

\begin{abstract}
Imaging genetics is a growing field that employs structural or functional neuroimaging techniques to study individuals with genetic risk variants potentially linked to specific illnesses. This area presents considerable challenges to statisticians due to the heterogeneous information and different data forms it involves. In addition, both imaging and genetic data are typically high-dimensional, creating a ``big data squared'' problem, Moreover, brain imaging data contains extensive spatial information. Simply vectorizing tensor images and treating voxels as independent features can lead to computational issues and disregard spatial structure.	
This paper presents a novel statistical method for imaging genetics modeling while addressing all these challenges. We explore a Canonical Correlation Analysis based linear model for the joint modeling of brain imaging, genetic information, and clinical phenotype, enabling the simultaneous detection of significant brain regions and selection of important genetic variants associated with the phenotype outcome. Scalable algorithms are developed to tackle the ``big data squared'' issue. We apply the proposed method to explore the reaction speed, an indicator of cognitive functions, and its associations with brain MRI and genetic factors using the UK Biobank database. Our study reveals a notable connection between the caudate nucleus region of brain and specific significant SNPs, along with their respective regulated genes, and the reaction speed.
\end{abstract}

\keywords{Neuroimaging Data \and Imaging genetics \and Cognitive functions \and Canonical correlation analysis \and Kronecker product}

\section{Introduction}

\begin{figure}[h!]
\begin{center}
\includegraphics[width=5in]{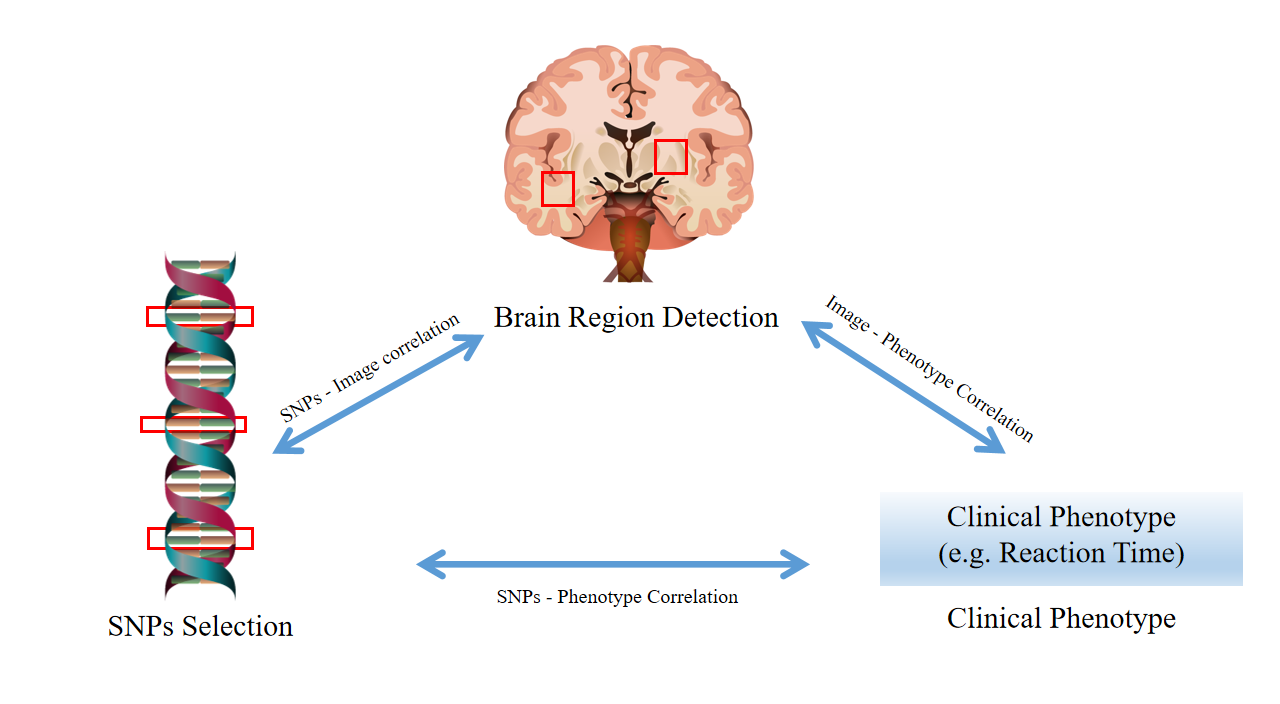}
\end{center}
\caption{An illustration of the connections between brain imaging, genetics information and phenotype outcome.
	 \label{fig1}}
\end{figure}

Imaging genetics is an emerging field that focuses on using structural or functional neuroimaging techniques to study individuals with genetic risk variants that possibly associated with certain illnesses. In particular, the diagnosis and treatment of many brain disorders are benefited tremendously by modern imaging technologies such as magnetic resonance imaging (MRI) or functional MRI (fMRI). This project concentrates on developing novel statistical methods to joint modeling brain imaging and genetics data, and establishing connections between these data with specific clinical phenotype. Figure \ref{fig1} provides an illustration of the objective of our analysis. 

The emergence of imaging genetics also poses significant challenges for statisticians. First, brain imaging, genetics information and phenotype outcomes contains heterogeneous information and are represented as different data forms, including (high-order) tensors, vectors and scalars. For instance,  a brain MRI scan yields a three order tensor, an fMRI scan generates a four order tensor, while genetics information (e.g., single nucleotide polymorphism, SNP) produces a vector.  Second, both imaging and genetic data are typically high dimensional, leading to a ``big data squared'' problem. Given that sample sizes are usually limited,
it is crucial to accurately locate brain regions of interest (ROI) associated with the phenotype outcomes and link these ROIs with available genetic information from SNPs in the subjects.
Third, brain imaging data encompasses abundant spatial information. Merely converting the tensor images into vectors would not only create computational issues, but also neglect the spatial structure and disrupt the signal region. Therefore, it is desired to develop statistical methods for imaging genetics analysis that can address these issues effectively. 
 
The initial approaches developed for analyzing imaging genetics data relied on substantial reductions to both imaging and genetic data. For instance, the analysis was limited to several summary measures (e.g., volume, thickness) of certain specific candidate ROIs in the brain and/or a particular set of genetic markers. Such type of analysis was first considered by \citet{stein2010voxelwise}, where each of 31,622 voxels was regressed on 448,293 SNPs separately across 740 elderly subjects. Moreover, rather than using the massive univariate approach, the voxel-wise approach has also be proposed in the literature, e.g., \citet{ge2012increasing,huang2015fvgwas}. These methods still involve fitting separate regression models for each brain location, but consider a collection of genetic markers at once instead of only a single marker. Furthermore, to account for the interdependence among various brain imaging phenotypes (summary measures), multivariate high-dimensional regression models with different matrix coefficients assumptions have been proposed, such as \citet{vounou2010discovering, wang2012identifying}. More recently, \citet{yu2022mapping} focused on Alzheimer’s Disease analysis and developed a two-step approach to map the genetic-imaging-clinical pathway.  \citet{zhao2023heart} examined MRI and genetic data from thousands of UK Biobank participants to investigate the associations between heart and brain health. 
 We acknowledge that the literature on imaging genetics discussed here is not exhaustive and refer to \citet{nathoo2019review} and \citet{zhu2023statistical} for comprehensive overviews.


It is important to emphasize that all the aforementioned approaches concentrate on the summarized measures of brain imaging, rather than the original data that represented as high order tensors. 
Consequently, a substantial amount of information may be lost. In the statistics community, various efforts have also been undertaken to develop methodologies for analyzing original brain imaging data.
A basic strategy involves first converting images into vectors and then considering vector-based methods.  Based on this strategy, various approaches have been proposed, including fused Lasso \citep{tibshirani2005sparsity} or other Total Variation  \citep{rudin1992nonlinear} based approaches, e.g., \citet{wang2017generalized}, Bayesian methods \citep{kang2018scalar,goldsmith2014smooth}, among others. While these approaches have proven effective in different applications, vectorizing images may not be an ideal strategy. Besides losing spatial information and breaking down signal regions, the generated ultra-high dimensional vectors may also encounter considerable computational challenges. 
Without converting images into vectors, \citet{zhou2013tensor} introduced a tensor regression framework for tensor image data. Their approach applies a canonical polyadic decomposition  (CPD) on the tensor coefficients, substantially decreasing the number of unknown parameters. \textcolor{black}{Apart from regression analysis, Canonical Correlation Analysis (CCA) has also been studied in the literature to model the relationships between blocks of variables. For instance, \citet{min2019tensor} introduced Tensor CCA (TCCA) to model the relationship between two blocks of tensor data and employed CPD to the canonical coefficients to reduce tensor dimension. Moreover, \citet{girka2024tensor} extended the two-block TCCA to Tensor Generalized CCA (TGCCA), allowing for the analysis of multiple data blocks.} 

In this paper, we develop a general framework for jointly modeling real brain MRI, genetics information and clinical phenotype outcomes. This proposed framework will enable us to identify brain ROIs associated with the phenotype outcomes and link these ROIs with available genetic information in the subjects.
Our approach is based on the multi-block canonical correlation analysis (mCCA), a powerful tool for exploring inter-relationships among heterogeneous datasets. Moreover, the proposed method achieves brain region detection by integrating a sparse Kronecker product decomposition (SKPD) framework \citet{wu2023sparse} on the canonical coefficients of image data. Under SKPD framework, we include a ``location indicator'' to locate significant brain regions and a ``dictionary'' to catch the shapes and intensities of the region. 
Consequently, we could not only preserves the spatial information inherent in brain MRI, but also efficiently reduces the dimension of imaging and genetic data.

We apply the proposed approach on real datasets from the UK Biobank (UKB). The results reveal a robust association between the caudate nucleus and cognitive function, as assessed by reaction time. Moreover, the credibility of these findings is ensured by 10 independent data subsets, thanks to the extensive sample size in UKB, and by corroborating with existing literature. Furthermore, we pinpoint several SNPs and their regulated genes that are involved in neurological disorders impacting cognitive function, with the majority being documented in prior studies and two being new discoveries in this research. More importantly, through a combined analysis of brain MRI images, genetic data, and phenotype variables, we could uncover potential genetic-image-clinical pathways, which are crucial for investigating the mechanisms underlying cognitive diseases.
	
The remaining of the paper is organized as follows. Section 2 provides a detailed explanation of the imaging, genetics, and phenotype data obtained from the UK Biobank study. In Section 3, we present the proposed method and its computation. Section 4 encompasses an in-depth analysis of the imaging genetics data in the UK Biobank. In Section 5, we conduct a comprehensive simulation study. Finally, Section 6 presents the concluding remarks.

\textcolor{black}{\textbf{Notation:} We use calligraphic letters, $\mathcal{A}$, $\mathcal{B}$, $\mathcal{C}$, to denote tensors. Bold upper-case letters, $\boldsymbol{A}$, $\boldsymbol{B}$, are used to denote matrices, and bold lower-case letters, $\boldsymbol{a}$, $\boldsymbol{b}$, to denote vectors. The vectorization operator is denoted as vec($\cdot$), and the inverse operator that transform a vector to its original matrix/tensor shape, is denoted as $\operatorname{vec}^{-1}(\cdot)$. The inner product is denoted by $\langle\cdot, \cdot\rangle$ and the Kronecker product is denoted by $\otimes$. For a vector $\boldsymbol{v}$, the $\ell_q$ norm is defined as $|\boldsymbol{v}|_q=\left(\sum_j\left|v_j\right|^q\right)^{1 / q}$. Additionally, $\boldsymbol{I}_n$ is used to denote an identity matrix of dimension $n \times n$.}

\section{UK Biobank}\label{sec:data}
In this study, we conduct an extensive analysis of the genetic information, T1-weighted structural brain MRI images and cognitive phenotypes data from the UKB database. This database has a large number of participants with multimodal imaging data acquired using homogeneous hardware and software. With the large number of participants in the UKB database, we have the opportunity to use the same database to discover and validate the findings independently, instead of adopting the ensemble methods to stabilize and validate the findings using the discovery set alone. Several studies of brain imaging based phenotypes have been reported \citep{Zhu2023science,Zhu2021science,li2021super,Bycroft2018} to take advantage of this large and rich database. However, a study of jointly modeling brain images, genetics information and clinical phenotype has not been conducted to the best of our knowledge. 

The primary aim of our study is to identify important brain regions in MRI imaging, as well as noteworthy SNPs in the genetic variants, that are associated with the speed of reaction, as quantified by the reaction time in the ``Snap'' card game (Field ID: 100103). The total number of subjects with all three datasets available is 42,770. To increase the reproducibility of results, we randomly partition the whole dataset into 10 batches for the purpose of internal validation. We report the commonly identified signals on multiple batches to ensure the reliability of the findings.
\subsection{Brain MRI}
We utilize T1-weighted structural brain MRI images in NIFTI format (Field ID: 20252) for our study. All participants underwent an imaging visit in 2014, and a subset of them were subsequently invited for a repeat imaging visit in 2019. For the purpose of our analysis, we specifically consider the brain MRI images obtained during the initial visit, while excluding those from the repeat visit. This selection process yielded a final sample size of MRI images as 42,770.

For each subject, the MRI images then undergo preprocessing: non-relevant regions are cropped, non-brain tissues are removed, and a gradient distortion correction is applied for bias correction. We employ two robust tools, BET \citep{smith} for brain extraction, converting brain masks into corrected input images, and FLIRT \citep{jenkin} for linear image registration using the standard space template MNI ICBM152 nonlinear 6th generation symmetric \citep{grab}. The images are then reoriented into the MNI152 space as a template and use FNIRT \citep{door} as nonlinear image registration. We get the final T1-weighted brain images that are projected into the MNI space, and the size of each image is $182\times218\times182$. To facilitate our analysis, we normalize each voxel in the images across all participants. In order to enhance analytical efficiency, we utilize the python package  \emph{SimpleITK}  to interpolate and resize the MRI images, resulting in a final size of $48 \times 60 \times 48$. 
\subsection{Genetic data}
The genetic data obtained from the UK Biobank comprises genotypes for 488,377 participants, which were assayed using two highly similar genotyping arrays. Specifically, a subset of 49,950 participants involved in the UK Biobank Lung Exome Variant Evaluation (UK BiLEVE) study underwent genotyping at 807,411 markers using the Applied Biosystems UK BiLEVE Axiom Array by Affymetrix. Subsequently, 438,427 participants were genotyped using the closely related Applied Biosystems UK Biobank Axiom Array consisting of 825,927 markers, where 95\% of the marker content are the same as in the UK BiLEVE Axiom Array. Further details regarding the genetic data can be found in \citet{Bycroft2018}. From this extensive pool, we select subjects whose brain MRI scan and relevant clinical phenotype data are all available.

To ensure the accuracy and integrity of genetic data, we use the \emph{plink} platform to conduct rigorous quality control procedures, which included screening out single-nucleotide polymorphisms (SNPs) that exceeded a 10\% missing rate among subjects, those with a minor allele frequency (MAF) less than 0.1, and SNPs with a p-value less than $10^{-5}$ in the Hardy-Weinberg equilibrium test. Having subjected the data to these quality control steps, we implement sure independence screening \citep{FanLv08} to further condense the dimension of SNPs data down to approximately 4,000. The finally obtained SNPs data for each subject are vectors with dimensions around 4,000 and with value 0, 1, 2 which are the counting number of minor allele for each SNP.

\subsection{Clinical phenotype}
The phenotype investigated in this study is the reaction speed, which is measured by the mean reaction time from the ``Snap'' card game. The mean reaction time is determined based on the  reaction times recorded over 12 rounds of the ``Snap'' card game. During each round, participants are presented with a pair of cards on a screen and instructed to press a button-box situated on the table in front of them as rapidly as possible, in the event that both cards are identical. The average reaction time is computed as the mean duration between the display of the cards and the initial click of the ``Snap'' button over the course of 12 rounds. It has been adopted to reflect the reaction time in several cognitive studies, e.g.,  \cite{kendall2019cognitive,lyall2022quantifying,davies2016genome, lyall2017associations}.

\section{Method}
\subsection{Problem Formulation and Method}
Suppose we observe $n$ i.i.d. samples, each with a three-order tensor imaging $\bmX_i\in \mathbb{R}^{D_1 \times D_2\times D_3}$  (e.g., MRI scan), and a genetic covariates $\bz_i\in \mathbb{R}^{q}$ (e.g., SNP), and a continuous  phenotype outcome $y_i\in \mathbb{R}$. We consider a high-dimensional setting where both image dimension $D_1\times D_2 \times D_3$ and genetic dimension $q$ could be larger than the sample size $n$. 
The objective is to develop a general framework to joint modeling the three data types, with an emphasis on identifying significant brain regions and selecting relevant genetic variants associated with the phenotype outcome.



Canonical correlation analysis (CCA) is a classical statistical method that employs cross-covariance matrices to derive information. The CCA was originally introduced for two vector sets, and later generalized to multi-block CCA to include $K$ vector sets with $K>2$. Different versions of mCCA have been proposed in the literature, such as SUMCOR \citep[Sum of Correlation,][]{van1984linear}, SSQCOR \citep[Sum of Squared Correlation,][]{hanafi2006analysis}, MAXVAR \citep[Maximum Variance,][]{kettenring1971canonical}. Here we shall focus on SUMCOR, which is defined as follows. Given $K$ sets of variables $\bx_{(j)}$ for $j=1,\ldots, K$, SUMCOR can be described as the following optimization problem
\bel{gcca}
\max_{[\ba_K]} && \sum_{j\neq k} \Cov(\bx_{(j)}\ba_j, \bx_{(k)}\ba_k), 
\cr  \text{s.t.} &&\Var (\bx_{(j)}\ba_j)=1, \ \ \  j=1\ldots, K.
\eel

In our analysis, brain imaging, genetics information along with the phenotype outcomes inherently form three groups of variables. The mCCA provides an natural way to build the connections among these three groups. Meanwhile, we note that the constraint in (\ref{gcca}) is better suited for imaging genetics analysis compared to an alternative constraint that imposes $\sum_{k=1}^K\Var  (\bx_{(j)}\ba_j)=1$. The alternative could result in extremely unbalanced constraints when considering such heterogeneous information.

As discussed before, a key step for imaging genetics analysis is to accurately locate significant brain regions and select relevant genetic variants associated with the phenotype outcome. Denote the target coefficients for genetics as $\btheta\in \mathbb{R}^{q}$. We assume that $\btheta$ is sparse, i.e., $\|\btheta\|_0\le s_0$, where $s_0\ll q$. In other words, there are at most $s_0$ genetic variants that are truly associated with the phenotype outcome.

To locate significant brain regions, we integrate mCCA with the sparse Kronecker product decomposition (SKPD) framework introduced in \citet{wu2023sparse}, which offers a convenient way for image region detection.
Denote the target coefficients tensor for imaging as $\bmC\in \mathbb{R}^{D_1 \times D_2\times D_3}$. 
We assume that 
$\bmC$ could be decomposed as 
\bel{skpd2}
\bmC=\sum_{r=1}^R\bmA_r\otimes\bmB_r, \ \ \|\bmA_r\|_{0} \le s_r,  \ \  \ r=1,2,\ldots, R,
\eel
where $\otimes$ is the Kronecker product, $\bmA_r$ and $\bmB_r$ are unknown tensors of size $p_1\times p_2\times p_3$ and $d_1\times d_2\times d_3$, respectively, and $\bmA_r$ are assumed to be sparse with certain sparsity levels $s_r$. The sizes of $\bmA_r$ and $\bmB_r$ are assumed unknown, but certainly need to satisfy $p_1d_1=D_1$, $p_2d_2=D_2$ and $p_3d_3=D_3$. 

When the sizes of $\mathcal{A}_r$ and $\mathcal{B}_r$ are unknown, the Sparse Kronecker product decomposition is not unique in general.
Nevertheless, when the size of $\mathcal{B}_r$ (or equivalently $\mathcal{A}_r$) is given, the decomposition can be identified with an orthogonality constraint. Thus, we  assume that
$\langle \bmA_r,\bmA_l \rangle=1 $ if $r=l$ and 0 otherwise.
In fact, the identification of SKPD can be reduced to that of sparse Singular Value Decomposition (SVD). We refer to \citet{wu2023sparse} for more discussion on the identifiability issue of SKPD.
.

In this formulation,  
the small block tensors $\bmB_r$ could be understood as the ``dictionaries'' and indicate the ``shape'' and ``intensity'' of the signal.
	While the tensors $\bmA_r$ are the ``location indicators'' to locate the signal blocks. For example, $\{\bmA_r\}_{j,k,l}\neq 0$ for certain $1\le j\le p_1$, $1\le k\le p_2$, $1\le l\le p_3$ and $1\le r\le R$ suggests that the region $\left[\left((j-1)d_1+1\right): j d_1;\  \left((k-1)d_2+1\right): k d_2; \ \left((l-1)d_3+1\right): k d_3\right]$ contains signal. 
 
 To estimate $\btheta$, $[\bmA_r]$ and $[\bmB_r]$, we consider the following regularized optimization problem:
	\bel{obj}
	\min_{[\bmA_r], [\bmB_r],\btheta} &&
		-
		\Big\{\Cov(\langle\bmX, \bmC\rangle,\langle \bz, \btheta\rangle)+\Cov(\langle\bmX, \bmC\rangle, y)+\Cov(\langle\bz, \btheta\rangle, y)\Big\}\cr &&+\lam_1\|\btheta\|_1+\lam_2  \sum_{r=1}^R\|\text{vec}(\mathcal{A}_r)\|_1\cr
		 \text{s.t.} &&
		 \Var(\langle\bz, \btheta\rangle)\le1, \ \ 
		 \Var(\langle\bmX, \bmC\rangle)\le 1,\ \ \bmC=\sum_{r=1}^{R}\mathcal{A}_r \otimes \mathcal{B}_r, 
	\eel
	where $\lam_1$ and $\lam_2$ are regularization parameters to account for the sparseness of $\btheta$ and $\bmA_r$. 
Note that we relax the equality constraints in (\ref{gcca}) to the inequality in (\ref{obj}) for computational considerations. Such kind of relaxations have also been considered in the literature on CCA, for example, \cite{chi2013imaging}.
On the other hand, we note that appropriate normalization is needed before implementing (\ref{obj}). In particular, we standardize the scalar outcomes $y$ such that $\Var(y)=1$. As a result, the coefficients estimation for $y$ would not be needed. 
When the brain imaging, genetics information and phenotype outcomes are properly centered, the sample version of (\ref{obj}) becomes the following form: 
	\bel{obj2}
\min_{[\bmA_r], [\bmB_r],\btheta} &&
-
\sum_{i=1}^n \Big(y_i+\langle\bmX_i, \bmC\rangle \Big) \bz_i^\T\btheta-y_i\langle\bmX_i, \bmC\rangle + \lam_1\|\btheta\|_1+\lam_2  \sum_{r=1}^R\|\text{vec}(\mathcal{A}_r)\|_1\cr
	 \text{s.t.} &&\sum_{i=1}^n (\bz_i^\T\btheta)^2\le n, \ \ 
\sum_{i=1}^n(\langle\bmX_i, \bmC\rangle)^2\le n,\ \ \bmC=\sum_{r=1}^{R}\mathcal{A}_r \otimes \mathcal{B}_r. 
\eel

\subsection{Computation with fixed dimension}\label{sec3-2}
In this subsection, we discuss the computation of the proposed minimization problem. We shall first consider the case that the block sizes of $(d_1,d_2,d_3)$, ranks $R$ and regularizations $(\lam_1, \lam_2)$ are given. The tuning parameter selection will be discussed in Section \ref{sec3-3}.

To solve (\ref{obj2}), a key step is to apply an tensor reshaping operator to the original images. Let $p=p_1p_2p_3$ and $d=d_1d_2d_3$. For any tensor $\bmC$ of dimension $D_1\times D_2\times D_3$, define $\mR: \mathbb{R}^{D_1\times D_2\times D_3}\rightarrow \mathbb{R}^{p\times d}$, as a mapping from tensor $\bmC$ to matrix
 \bel{r}
 	\mathcal{R}(\mathcal{C})=\left[\operatorname{vec}\left(\mathcal{C}_{1,1,1}^{d_1, d_2, d_3}\right), \ldots, \operatorname{vec}\left(\mathcal{C}_{1,1, p_3}^{d_1, d_2, d_3}\right), \ldots, \operatorname{vec}\left(\mathcal{C}_{1, p_2, 1}^{d_1, d_2, d_3}\right), \ldots, \operatorname{vec}\left(\mathcal{C}_{1, p_2, p_3}^{d_1, d_2, d_3}\right), \ldots,\right. \cr \left.\operatorname{vec}\left(\mathcal{C}_{p_1, 1,1}^{d_1, d_2, d_3}\right), \ldots, \operatorname{vec}\left(\mathcal{C}_{p_1, 1, p_3}^{d_1, d_2, d_3}\right), \ldots, \operatorname{vec}\left(\mathcal{C}_{p_1, p_2, 1}^{d_1, d_2, d_3}\right), \ldots, \operatorname{vec}\left(\mathcal{C}_{p_1, p_2, p_3}^{d_1, d_2, d_3}\right)\right]^{\top}
 \eel
 where $\bmC_{j,k,l}^{d_1,d_2,d_3}$ is the $(j,k,l)$-th block of $\bmC$ of size $d_1\times d_2\times d_3$. This operation essentially converts each block into a vector and form a new matrix with each row being the obtained vector.
 A desirable property of the operator $\mR(\cdot)$ is that when applied to Kronecker product $\bmA\otimes \bmB$, it holds that
 \bel{Rmat2}
 \mR(\bmA\otimes \bmB)=\tvec(\bmA)[\tvec(\bmB)]^\T.
 \eel
 The reshaping operator $\mR$ is crucial to the computation as it allows us to transform the original tensor operations into matrix analyses. 
 This is significant as it circumvents the complex tensor decompositions, such as canonical polyadic decomposition or the Tucker decomposition, and unifies the tensor analysis into matrix analysis problem.
 
 By letting $\tbX_i = \mathcal{R}(\bmX_i)$, $\balpha_r=\tvec(\bmA_r)$ and $\bbeta_r=\tvec(\bmB_r)$, the problem (\ref{obj2}) could be transformed into the following minimization:
 \bel{obj3}
\min_{[\ba_r], [\bb_r],\btheta} &&
-
\sum_{i=1}^n \Big(y_i+\bz_i^\T\btheta\Big)\Big(\sum_{r=1}^R\balpha_r^\T\tbX_i\bbeta_r \Big) -y_i\bz_i^\T\btheta+ \lam_1\|\btheta\|_1+\lam_2  \sum_{r=1}^R\|\balpha_r\|_1\cr
\text{s.t.} &&\sum_{i=1}^n (\bz_i^\T\btheta)^2\le n, \ \ 
\sum_{i=1}^n\left(\sum_{r=1}^R\balpha_r^\T\tbX_i\bbeta_r\right)^2 \le n.
\eel

We apply an alternating minimization algorithm to solve  (\ref{obj3}). Let $\balpha=[\balpha_1,\ldots,\balpha_R]\in\mathbb{R}^{(p_1p_2p_3)\times R}$ and $\bbeta=[\bbeta_1,\ldots,\bbeta_R]\in\mathbb{R}^{(d_1d_2d_3)\times R}$. Then $\balpha$, $\bbeta$ and $\btheta$ could be updated in an alternating fashion. 

We start with both $\hbalpha^{(0)}$ and $\hbbeta^{(0)}$ being all zero matrices. 
Then given $[\hbalpha]$ and $[\hbbeta]$, some linear algebra suggests that $\hbtheta$ can be obtained by
$\hbtheta=\|\hbSigma_1^{1/2}\tbtheta\|_2^{-1} \tbtheta$ if $\|\hbSigma_1^{1/2}\tbtheta\|_2>0$ and $0$ otherwise, where $\hbSigma_1=(1/n)\sum_{i=1}^n \bz_i\bz_i^\T$ is the estimated sample covariance matrix for $\bz$, 
and $\tbtheta$ is the solution to be the following Lasso problem:
\bel{theta}
\tbtheta =\argmin_{\btheta} \frac{1}{2}\left\|\hbSigma_1^{-1/2}\bZ^\T\bGamma_1 - \hbSigma_1^{-1/2}\btheta\right\|_2^2 +\lam_1 \|\btheta\|_1.
\eel
Here $\bZ=(\bz_1,\ldots,\bz_n)^\T\in\mathbb{R}^{n\times q}$ and $\bGamma_1\in\mathbb{R}^n$ with the $i$-th element being
$\{\Gamma_1\}_i=\sum_{r=1}^R\hbalpha_r^\T\tbX_i\hbbeta_r +y_i$.  On the other hand, we shall note that the calculation of $\hbSigma_1^{-1/2}$ is needed in (\ref{theta}). While in a high-dimensional setting where $q>n$,  $\hbSigma_1$ is not invertable. Under such a scenario, we may modify the estimate of $\hbSigma_1$ to ensure its invertability, for instance,  $\hbSigma_1=(1/n)\sum_{i=1}^n \bz_i\bz_i^\T+\tau \bI_{q}$, with $\tau$ being a small constant.

Similarly, $\hbalpha$ could also be obtained by a Lasso problem given $\hbtheta$ and $\hbbeta$. Let $\{\bX_{\beta}\}_i=\tvec(\tbX_i\hbbeta)\in \mathbb{R}^{R(p_1p_2p_3)}$ for $i=1,\ldots, n$ and $\bX_{\beta}=\left(\{\bX_{\beta}\}_1,\ldots, \{\bX_{\beta}\}_n\right)^\T\in \mathbb{R}^{n\times R(p_1p_2p_3)}$ be the combined matrix. Further let 
$\hbSigma_2=(1/n)\sum_{i=1}^n \{\bX_{\beta}\}_i\{\bX_{\beta}\}_i^\T+\tau\boldsymbol{I}_{Rp_1p_2p_3}$ be the regularized sample covariance matrix for $\bX_{\beta}$.
Then  $\hbalpha$ can be obtained by 
 $\hbalpha=\tvec^{-1}\left(\|\hbSigma_2^{1/2}\tbalpha\|_2^{-1} \tbalpha\right)$, where $\tbalpha$ is the solution to the following Lasso problem:
\bel{alpha}
\tbalpha =\argmin_{\balpha} \frac{1}{2}\left\| \hbSigma_2^{-1/2}\bX_{\beta}^{\T}\bLambda - \hbSigma_2^{-1/2}\balpha\right\|_2^2 +\lam_2 \|\balpha\|_1 .
\eel
Here  $\bLambda \in \mathbb{R}^{n}$ and its $i$-th element is $\Lambda_i = \hbtheta^{\T}z_i + y_i$. To ensure the orthogonality across $\tbalpha$, we further update $\tbalpha$ by $\boldsymbol{\hat{\alpha}}^{(t+1)}\leftarrow \boldsymbol{\tilde{\alpha}}^{(t+1)} [(\boldsymbol{\tilde{\alpha}}^{(t+1)})^{T}\boldsymbol{\tilde{\alpha}}^{(t+1)}]^{-1/2} $. 
when $(\boldsymbol{\hat{\alpha}}^{(t+1)})^{T}\boldsymbol{\hat{\alpha}}^{(t+1)}$ is singular, we add a regularization parameter $\tau$ to avoid singularity and update $\hat{\alpha}$ by $\boldsymbol{\hat{\alpha}}^{(t+1)}\leftarrow \boldsymbol{\tilde{\alpha}}^{(t+1)} [(\boldsymbol{\tilde{\alpha}}^{(t+1)})^{T}\boldsymbol{\tilde{\alpha}}^{(t+1)}+\tau I_R]^{-1/2}$. The value of $\tau$ is fixed to be $\tau=0.01$ throughout the numerical studies in this paper.

	

Finally, given $\hbtheta$ and $\hbalpha$, $\hbeta$ could  be solved by an ordinary least squares (OLS) problem. Let $\{\bX_{\alpha}\}_i=\tvec(\tbX_i\halpha)\in \mathbb{R}^{R(d_1d_2d_3)}$ for $i=1,\ldots, n$ and $\bX_{\alpha}=\left(\{\bX_{\alpha}\}_1,\ldots, \{\bX_{\alpha}\}_n\right)^\T\in \mathbb{R}^{n\times R(d_1d_2d_3)}$. Further let 
$\hbSigma_3=(1/n)\sum_{i=1}^n \{\bX_{\alpha}\}_i\{\bX_{\alpha}\}_i^\T + \tau\boldsymbol{I}_{Rd_1d_2d_3}$ be the regularized sample covariance matrix for $\bX_{\alpha}$.
Then  $\hbbeta$ can be obtained by 
$\hbbeta=\tvec^{-1}\left(\|\hbSigma_3^{1/2}\tbbeta\|_2^{-1} \tbbeta\right)$, where $\tbbeta$ is the solution to the following OLS problem:
\bel{beta}
\tbbeta =\argmin_{\bbeta} \frac{1}{2}\left\| \hbSigma_3^{-1/2}\bX_{\alpha}^{\T}\bLambda - \hbSigma_3^{-1/2}\bbeta\right\|_2^2 .
\eel
\begin{algorithm}[H]
\caption{Alternating Minimization Algorithm}\label{algorithm}
\KwIn{$\mathcal{X}_i$,$\bz_i$, $y_i$ for $i=1, \ldots, n$ and $\lam_1$, $\lam_2$, $R$, $d_1$, $d_2$, $d_3$. } 
Initialization: Initialize $\hbalpha^{(0)}$ and $\hbbeta^{(0)}$ to matrices with all entries equal to 1. Initialize sample covariance matrix of $z$ as $\hbSigma_1\leftarrow (1/n)\sum_{i=1}^n \bz_i\bz_i^\T+\tau \bI_{q}$\;
\For{t in $0,1,2, \ldots T-1$}{
	 $\{\Gamma_1\}_i\leftarrow \sum_{r=1}^R\hbalpha_r^{(t)\T}\tbX_i\hbbeta_r^{(t)} +y_i$ \;$\tbtheta^{(t+1)} \leftarrow \argmin_{\btheta} \frac{1}{2}\left\|\hbSigma_1^{-1/2}\bZ^\T\bGamma - \hbSigma_1^{1/2}\btheta\right\|_2^2 +\lam_1 \|\btheta\|_1$\;
$\hbtheta^{(t+1)}\leftarrow \|\hbSigma_1^{1/2}\tbtheta^{(t+1)}\|_2^{-1} \tbtheta^{(t+1)}$\;
$\hbSigma_2\leftarrow (1/n)\sum_{i=1}^n \{\bX_{\beta}\}_i\{\bX_{\beta}\}_i^\T + \tau\boldsymbol{I}_{Rp_1p_2p_3}$, where $\{\bX_{\beta}\}_i=\tvec(\tbX_i\hbbeta^{(t)})$\; $\Lambda_i \leftarrow  \hbtheta^{(t+1)\T}z_i + y_i$  \;
$\tbalpha^{(t+1)} \leftarrow \argmin_{\balpha} \frac{1}{2}\left\| \hbSigma_2^{-1/2}\bX_{\beta}^{\T}\bLambda - \hbSigma_2^{-1/2}\balpha\right\|_2^2 +\lam_2 \|\balpha\|_1$\;
$\hbalpha^{(t+1)}\leftarrow \tvec^{-1}\left(\|{\hbSigma_{2}^{1/2}\tbalpha^{(t+1)}\|_2^{-1} \tbalpha^{(t+1)}}\right)$\;
$\hbalpha^{(t+1)}\leftarrow \hbalpha^{(t+1)} [(\hbalpha^{(t+1)})^{\T}\hbalpha^{(t+1)}]^{-1/2} $\;
 $\hbSigma_3\leftarrow (1/n)\sum_{i=1}^n \{\bX_{\alpha}\}_i\{\bX_{\alpha}\}_i^\T + \tau\boldsymbol{I}_{Rd_1d_2d_3}$, where $\{\bX_{\alpha}\}_i=\tvec(\tbX_i\hbalpha^{(t+1)})$ \;
 $\tbbeta^{(t+1)} =\argmin_{\bbeta} \frac{1}{2}\left\| \hbSigma_3^{-1/2}\bX_{\alpha}^{\T}\bLambda - \hbSigma_3^{-1/2}\bbeta\right\|_2^2$ \; $\hbbeta^{(t+1)}\leftarrow\tvec^{-1}\left(\|\hbSigma_3^{1/2}\tbbeta^{(t+1)}\|_2^{-1} \tbbeta^{(t+1)}\right)$ }
\Return $\hbtheta^{(T)}$, $\hbalpha^{(T)}$, $\hbbeta^{(T)}$
\end{algorithm}

\subsection{Convergence Analysis}\label{sec3-4}

In this subsection, we provide a brief discussion on the convergence properties of the proposed alternating minimization algorithm. 

We shall first recognize that optimization problem (\ref{obj3}) is nonconvex, and the convergence properties of the alternating minimization algorithm under a nonconvex objective function are generally challenging to establish. However, our problem falls into the category of multiconvex optimization problems, where the feasible set and objective function are nonconvex but convex within each block of variables.
Moreover, the proposed alternating minimization algorithm can be viewed as a Gauss-Seidel type block coordinate descent (BCD) method. Particularly, the three blocks of variables, $\balpha$, $\bbeta$, and $\btheta$, are updated cyclically, with each block being optimized while keeping the other blocks fixed at their most recently updated values. The convergence properties of Gauss-Seidel type BCD for multiconvex problems have been studied in the literature.
For instance, \citet{xu2013block} investigated regularized block multiconvex optimization problems, established global convergence, and estimated the asymptotic convergence rate for the Gauss-Seidel type BCD algorithm under the Kurdyka–Lojasiewicz  property. By applying the results of \citet{xu2013block}, the global convergence of our alternating minimization algorithm can also be ensured, given that a good initialization is suggested.
	Additionally, \citet{tseng2001convergence} provided a convergence analysis for a block coordinate descent algorithm applied to a non-differentiable objective function with certain separability and regularity properties. More recently, \citet{lange2021nonconvex} presented a comprehensive survey on the convergence of the Majorization-Minimization (MM) algorithm, which includes special cases such as the expectation-maximization algorithm, proximal gradient algorithm, concave-convex procedure, and more.
	Interested readers are referred to the aforementioned papers for a deeper understanding of the convergence properties of general alternating minimization algorithm.

\subsection{Tuning parameter selection}\label{sec3-3}

The tuning parameter involved in the optimization problem includes: image block sizes $(d_1,d_2,d_3)$, ranks $R$ and regularizations $(\lam_1, \lam_2)$. 
Note that determining the block sizes $(d_1,d_2,d_3)$ is equivalent to selecting the sizes of location indicator matrices $(p_1,p_2,p_3)$.
	Here, we follow the strategy of \citet{wu2023sparse} and treat the block sizes $(d_1,d_2,d_3)$ as fixed. The major reason behind is that as long as the rank $R$ is large enough, the sparse Kronecker product decomposition (\ref{skpd2}) could approximate arbitrary signal shapes even with fixed  $(d_1,d_2,d_3)$. As a result, meticulously adjusting $(d_1,d_2,d_3)$ may not provide substantial benefits to the optimization problem; instead, it could introduce considerable computational burdens.
	
	
	Given fixed block sizes, the best ranks $R$ and regularizations $(\lam_1, \lam_2)$ are obtained by minimizing the following modified Bayesian Information Criterion (BIC)
 \bel{obj5}
-
\frac{1}{n}\sum_{i=1}^n \Big(y_i+\bz_i^\T\hbtheta\Big)\Big(\sum_{r=1}^R\hbalpha_r^\T\tbX_i\hbbeta_r \Big) -y_i\bz_i^\T\hbtheta+ \frac{\log n}{n}\|\hbtheta\|_1+ \frac{\log n}{n}  \sum_{r=1}^R\|\hbalpha_r\|_1
\eel
where $\hbtheta=\hbtheta(\lam_1,\lam_2, R)$, $\hbalpha_r=\hbalpha_r(\lam_1,\lam_2, R)$ and $\hbbeta_r=\hbbeta_r(\lam_1,\lam_2, R)$ are the estimated values given $\lam_1$, $\lam_2$, and  $R$.

\section{The UK Biobank Analysis}
As discussed in Section \ref{sec:data}, we divide the entire dataset into 10 separate batches, each containing approximately 4,000 samples. The MRI images and SNPs are first demeaned, while the phenotype outcome mean reaction time is standardized prior to analysis.

We implement proposed approach with a block size $(d_1,d_2,d_3)=(4,5,4)$, while the resulting ``location indicator'' is of size $(p_1,p_2,p_3)=(12,12,12)$. The hyper-parameters  $\lambda_1$, $\lambda_2$ and $R$ are selected using the BIC described in (\ref{obj5}). We also implemented a special version of our method with the Kronecker rank fixed to be $R=1$. This streamlined model can be regarded as a reference point. Additionally, we note that if we consider voxel-wise blocks, i.e., $(d_1,d_2,d_3)=(1,1,1)$, the proposed approach reduces to the classic Sparse Canonical Correlation Analysis (SCCA). The SCCA is implemented as a competitor in Section \ref{sec:sims} on simulation studies. However, when analyzing UKB database, SCCA is not computationally viable due to the high-dimensional vector that results from flattening the brain MRI tensor image. Specifically, SCCA requires the processing and storage of matrices with dimensions of $138,240 \times 138,240$, which is impractical with conventional computational resources. Indeed, there is a shortage of existing methods capable of simultaneously analyzing high-dimensional image and genetic data, as well as phenotypic outcomes. As a result, we chose to employ a technique that separately examines ``phenotype vs imaging'' and ``phenotype vs genetics''. We apply the Tensor Regression model with Lasso regularization (TR Lasso) proposed by \cite{zhou2013tensor} for assessing ``phenotype vs imaging'', while employing a standard Lasso for ``phenotype vs genetics''.



\begin{figure}[h!]
	\centering
	\includegraphics[scale=0.5]{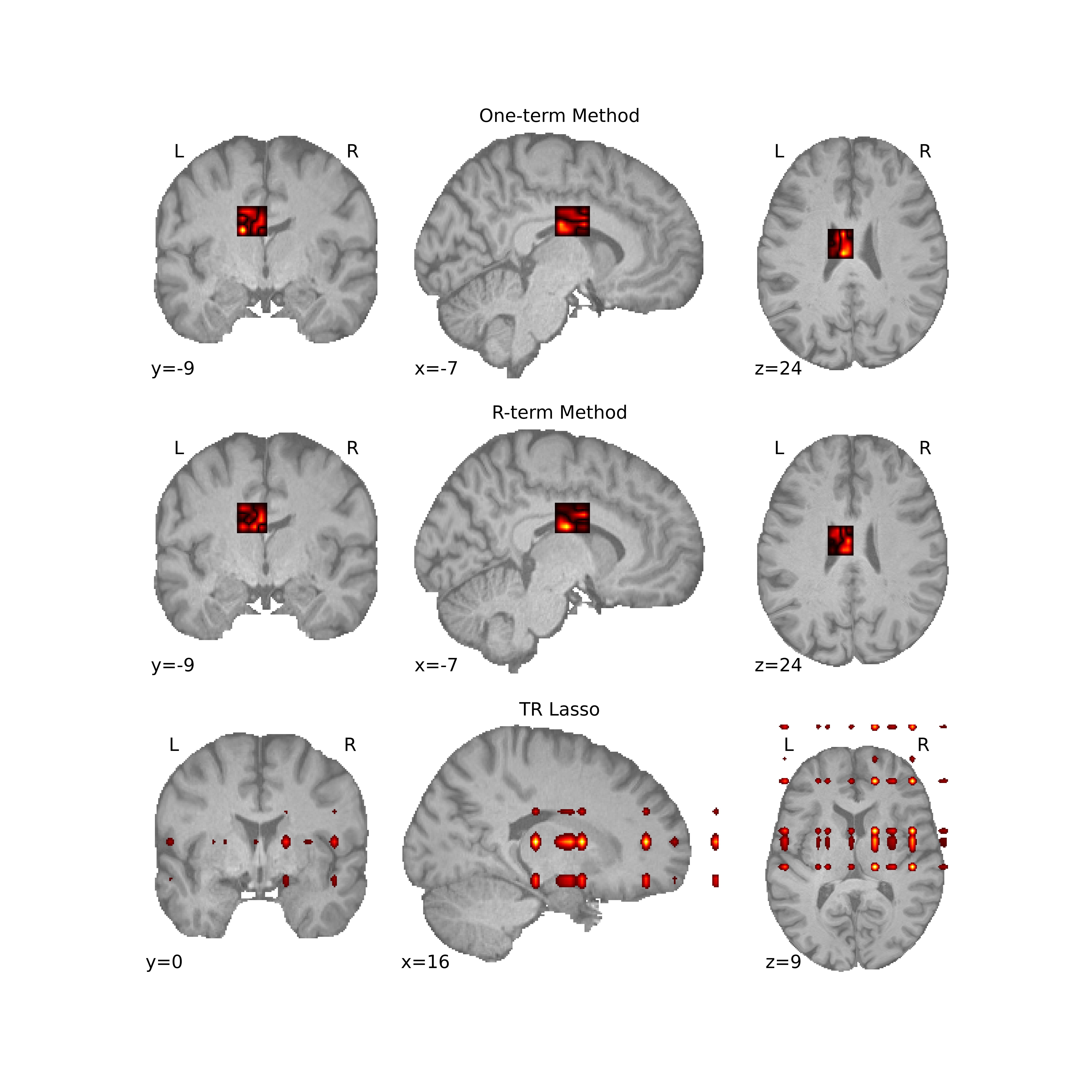}
	\vspace{-0.5in}
	\caption{The estimated tensor coefficients for UK Biobank Brain MRI data.}
	\label{fig2}
\end{figure}

\begin{figure}[h!]
	\centering
	\includegraphics[scale=0.25]{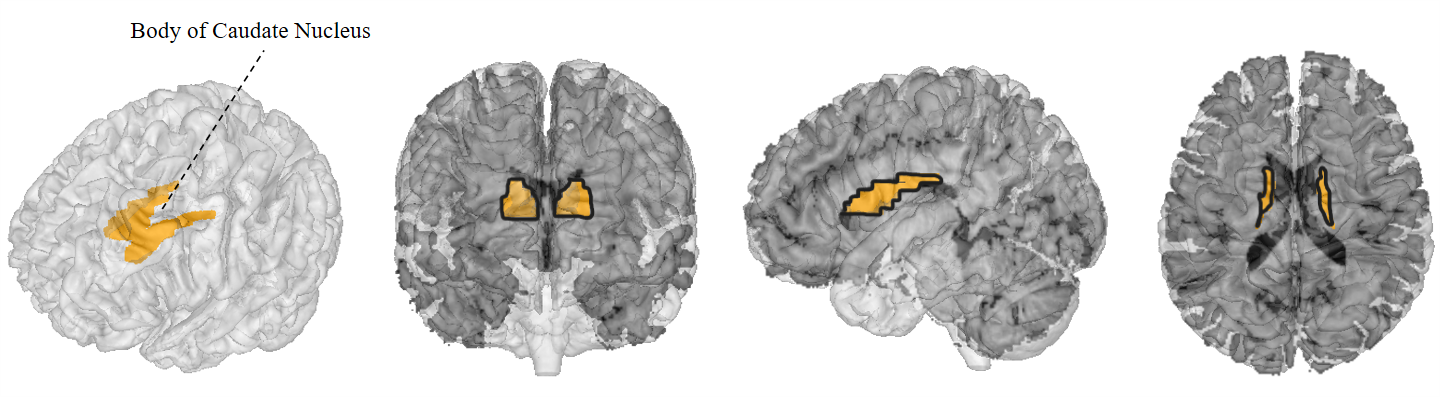}
	\caption{The 3-D (1st) and 2-D slice (2nd to 4th)  visualization of actual locations of caudate nucleus region.}
	\label{fig3}
\end{figure}

Our findings conclusively show that the regions identified by both one-term and R-term methods significantly overlap with the main body of the caudate nucleus. Specifically, the same region is identified in 9 out of the 10 batches. Conversely, the regions pinpointed by the TR Lasso method appear to be sporadic and less consistent. This further emphasizes the reliability and robustness of the proposed method. The tensor coefficients estimated from one batch are illustrated in Figure \ref{fig2}.  In Figure \ref{fig3}, the selected region, specifically the body of the caudate nucleus, is visually represented in the 3D brain MRI image with its 2-D slice sections.

The caudate nucleus, an integral part of the basal ganglia, is composed of two caudate nuclei located near the thalamus at the center of the brain. It plays a significant role in various functions, including planning and executing movement, learning, memory, reward, motivation, emotion, and romantic interaction \citep{driscoll2020neuroanatomy}. Numerous studies have established a link between the caudate nucleus and several significant neurological disorders, such as Alzheimer's disease \citep{jiji2013}, Parkinson's disease \citep{grahn2009}, Attention-deficit hyperactivity disorder \citep{schrimsher2002}, and Schizophrenia \citep{takase2004}.

The phenotype variable in this study is the reaction time, assessed through the ``Snap'' game. This game measures both the reaction speed and motor function of the participants, as they are required to press a button as quickly as possible when they see a pair of identical cards on the screen. This task is closely related to the caudate nucleus's role in processing visual information, cognitive function, and movement control \citep{driscoll2020neuroanatomy}. Emerging evidence suggests that the caudate nucleus is instrumental in goal-directed behavior, with studies on Parkinson's disease and the increase in blood flow in the left caudate nucleus during complex planning tasks providing direct evidence for this claim.

Functional MRI (fMRI) studies have also confirmed the caudate's involvement in various executive processes, particularly tasks requiring critical processes for goal-directed action, which robustly activate the caudate. For instance, in studies by \cite{knutson2001anticipation,knutson2001dissociation}, the outcome (obtaining reward or avoiding punishment) is determined by the subject's response speed in a target detection task, similar to the reaction time in our study. These studies generally find caudate activity when the subjects' decisions or reaction time determine the outcome. However, when responses do not determine the outcome (or only non-effortful responses are necessary), the caudate does not respond. This suggests that the caudate is sensitive to action reinforcement rather than rewards themselves. For more details, we recommend an extensive review \citep{grahn2008cognitive}. Our findings are in line with these existing studies, further reinforcing the established understanding of the caudate nucleus's role in goal-directed actions.

\begin{figure}[h!]
	\centering
	\includegraphics[scale=0.40]{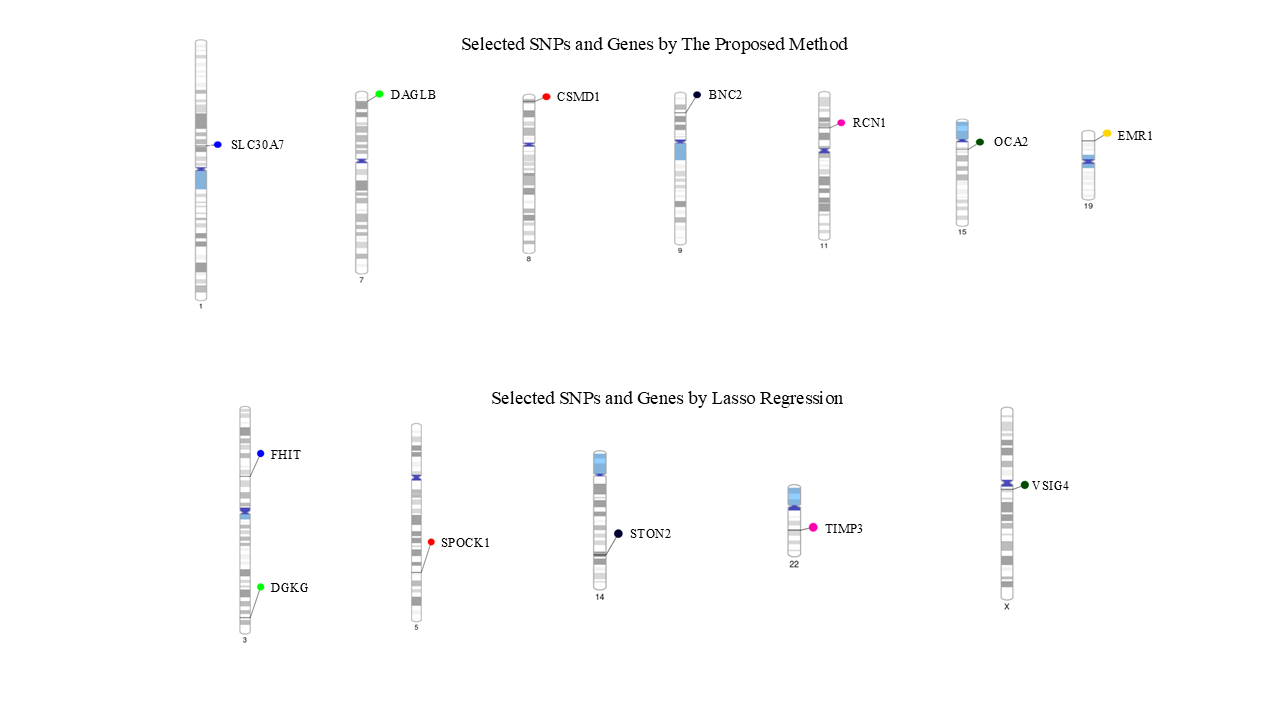}
	\caption{The locations of selected SNPs and genes on the chromosomes by our proposed method and Lasso regression.}
	\label{fig4}
\end{figure}

Beside brain region detection in MRI, our method also incorporates the simultaneous selection of important SNPs. Recall that we additionally implemented Lasso regression for ``phenotype vs genetics'' without incorporating imaging information. In Table \ref{tab1}, we display the SNPs with regulated genes that are consistently identified in three or more batches using our proposed method, while Table \ref{tab2} presents the results from the Lasso regression. We document the literature and studies that report on brain-related, psychiatric, or cognitive traits. The locations of the selected genes on the chromosomes are illustrated in Figure \ref{fig4}.

\begin{table}[h!]
	\centering
    \fontsize{11pt}{11pt}\selectfont
	\begin{tabular}{cccc}
		SNP & Chr & Genes &Reported brain-related/pyschiatric/cognitive trait(s) \\
		\hline 
		rs6693456 & 1 & SLC30A7 & \makecell{Multiple sclerosis \\ \citep{MS2011,MS2019}} \bigskip \\
		\bigskip
		rs13240753 & 7 & DAGLB & Parkinsonism \citep{Liu2022}\\
		rs1595467 & 8 & CSMD1 & Schizophrenia \citep{trubetskoy2022} \\
        & &  & Reaction time measurement \citep{davies2018} \\
        & &  & Neuroticism measurement \citep{luciano2018} \bigskip \\
		rs11789463 & 9 & BNC2 & Unipolar depression \citep{cai2020minimal}\\
        & & & Attention deficit hyperactivity disorder \citep{rovira2020} \bigskip  \\
		rs2440251 & 11 & RCN1 & N/A \bigskip \\
		rs11637518 & 15 & OCA2 & Parkinson's disease \citep{latourelle2009}\\
        & & & Attention deficit hyperactivity disorder \citep{hawi2018case} \bigskip  \\
		rs388711 & 19 & EMR1 & N/A\\
	\end{tabular}
    \vspace{0.1in}
	\caption{Associated SNPs identified by proposed approach.}
	\label{tab1}
\end{table}

\begin{table}[h!]
	\centering
    \fontsize{11pt}{11pt}\selectfont
	\begin{tabular}{cccc}
		 SNP & Chr & Regulated Genes &Reported brain-related/pyschiatric/cognitive trait(s) \\
		\hline 
		rs73097582 & 3 & FHIT & \makecell{Unipolar depression \\ \citep{giannakopoulou2021}} \bigskip \\
		rs2284842 & 3 &  DGKG & \makecell{Unipolar depression \\ \citep{giannakopoulou2021}}\bigskip \\
		rs6596395 & 5 & SPOCK1 & \makecell{Cerebellum white matter volume change \\ \citep{brouwer2022}}\bigskip \\
		rs9323700 & 14 & STON2 & Schizophrenia \citep{Luan2011} \bigskip \\
		rs5749529 & 22 & TIMP3 & N/A \bigskip\\
		rs678475 & X & VSIG4 & N/A\\
		
	\end{tabular}
    \vspace{0.1in}
	\caption{Associated SNPs identified by Lasso Regression.}
	\label{tab2}
\end{table}

According to Table \ref{tab1}, most detected biomarkers are well-studied and associated with several cognitive functions and disorders, which are also connected with the caudate nucleus as previously discussed. Another significant advantage of our proposed method, which conducts a combined analysis of brain MRI images, genetic data, and phenotype variables, is its potential to uncover a genetic-image-clinical pathway. This pathway holds scientific significance in exploring the mechanisms of various diseases.

Specifically, we identify the SNP \emph{rs13240753} in the \emph{DAGLB} gene. This gene functions as the primary 2-arachidonoyl-glycerol (2-AG) synthase in both human and mouse substantia nigra (SN) dopaminergic neurons (DANs). The levels of 2-AG can significantly influence dopamine release and DANs activity \citep{Liu2022}. Previous studies have established a link between the degree of dopaminergic neuron loss projecting to the caudate nucleus and Parkinson's disease \citep{grahn2008}. The caudate nucleus is the recipient of the dopamine released from the SN. As a result, a genetic-image-clinical pathway for Parkinson's disease becomes apparent: the \emph{DAGLB} gene's impact on dopamine release and DANs activity within the SN, which ultimately affects the dopaminergic neurons connected to the caudate nucleus. Changes in these neurons projected onto the caudate nucleus are closely linked to the pathogenesis of Parkinson's disease, thereby influencing overall cognitive function.

Another potential genetic-image-clinical pathway is associated with schizophrenia. We identify the SNP \emph{rs1595467} located on the \emph{CSMD1} gene. This gene has been linked to brain white matter volume through statistical analysis of polygenic risk scores \citep{oertel2015schizophrenia}. Furthermore, \citet{takase2004} reported a significant decrease in white matter volume within the caudate nucleus among individuals diagnosed with schizophrenia compared to healthy individuals. By integrating these findings, we can reasonably suggest a potential pathway for schizophrenia: a mutation in the \emph{CSMD1} gene leads to changes in white matter volume within the brain, which in turn contributes to various neurological disorders, including schizophrenia, ultimately affecting cognitive function.


However, among the SNPs detected by Lasso Regression, only one SNP, rs6596395, has a similar functional relevance to the caudate nucleus brain region, as previously discussed. Specifically, the SNP \emph{rs6596395} and its regulated gene, \emph{STON2}, have been identified to have a correlation with schizophrenia \citep{Luan2011}. This result highlights the inherent limitation of Lasso regression, as it solely focuses on the relationship between phenotypic and genetic information, thereby failing to incorporate the necessary components and understand cognitive functions in a comprehensive way. 

In conclusion, our results not only uncover the strong association between the caudate nucleus and the cognitive function measured by reaction time, but also reveal potential connections between identified biomarkers, the caudate nucleus and cognitive functions such as Parkinsonism, reaction time and attention deficit hyperactivity disorder. Compared to the vast number of genome-wide association studies, the studies on  genetic-image-clinical pathway are much more limited. Therefore, our novel findings shed light on medical studies aiming to decipher the mechanisms of cognitive disorders and the development of treatments.

\section{Simulation Studies}\label{sec:sims}
In this section, we conduct an extensive simulation analysis to demonstrate the performance of proposed method in image region detection and feature selection under various settings. 

We consider $N=1000$ subjects and each with an image $\bX_i$ of size $32\times 32$, genetic covariates $\bz_i$ of dimension $100$, and a scalar phenotype outcome $y_i$. The data sets are generated as follows. We first generate (vectorized) images and genetics covariates from a multivariate normal distribution with mean $\bf{0}$, covariance matrices $\bSigma_x\in\mathbb{R}^{32^2\times 32^2}$, $\bSigma_z\in\mathbb{R}^{100\times 100}$, and  
\bes
\bSigma_{xz}=\rho_1\boldsymbol{\Sigma}_{x}\tvec(\bC)\btheta^\T\bSigma_z \in\mathbb{R}^{32^2\times 100}.
\ees
Such a design would guarantee the correlations between $\langle\bX_i, \bC\rangle$ and $\langle\bz_i, \btheta\rangle$ to be $\rho_1$. Furthermore, we generate scalars  $y_i$ from normal distribution such that the correlation between $y_i$ and $\langle\bX_i, \bC\rangle$ equals to $\rho_2$. Detailed generation mechanisms can be found in the supplementary material.

We vary the correlations $(\rho_1, \rho_2)$, the covariance matrices $\bSigma_x$, the coefficients $\bC$ and examine their impacts to the proposed method. Specifically, we first consider different combinations of $\rho_1$ and $\rho_2$, both of which range from 0.5 to 0.8. 
Second, we study two types of covariance matrices $\bSigma_x$ and $\bSigma_z$, including identity and Toeplitz type matrices. For the identity case, we let $(\bSigma_x,\bSigma_z)=(\bI_{32^2},\bI_{100})$. While for the Toeplitz case, we let
$\bSigma_x(j,k)=0.9^{|j-k|}$ for $j,k=1,\ldots, 32^2$ and $\bSigma_z(j',k')=0.9^{|j'-k'|}$ for $ j',k'=1,\ldots, 100$. 

\begin{figure}[h!]
	\includegraphics[width=1\linewidth]{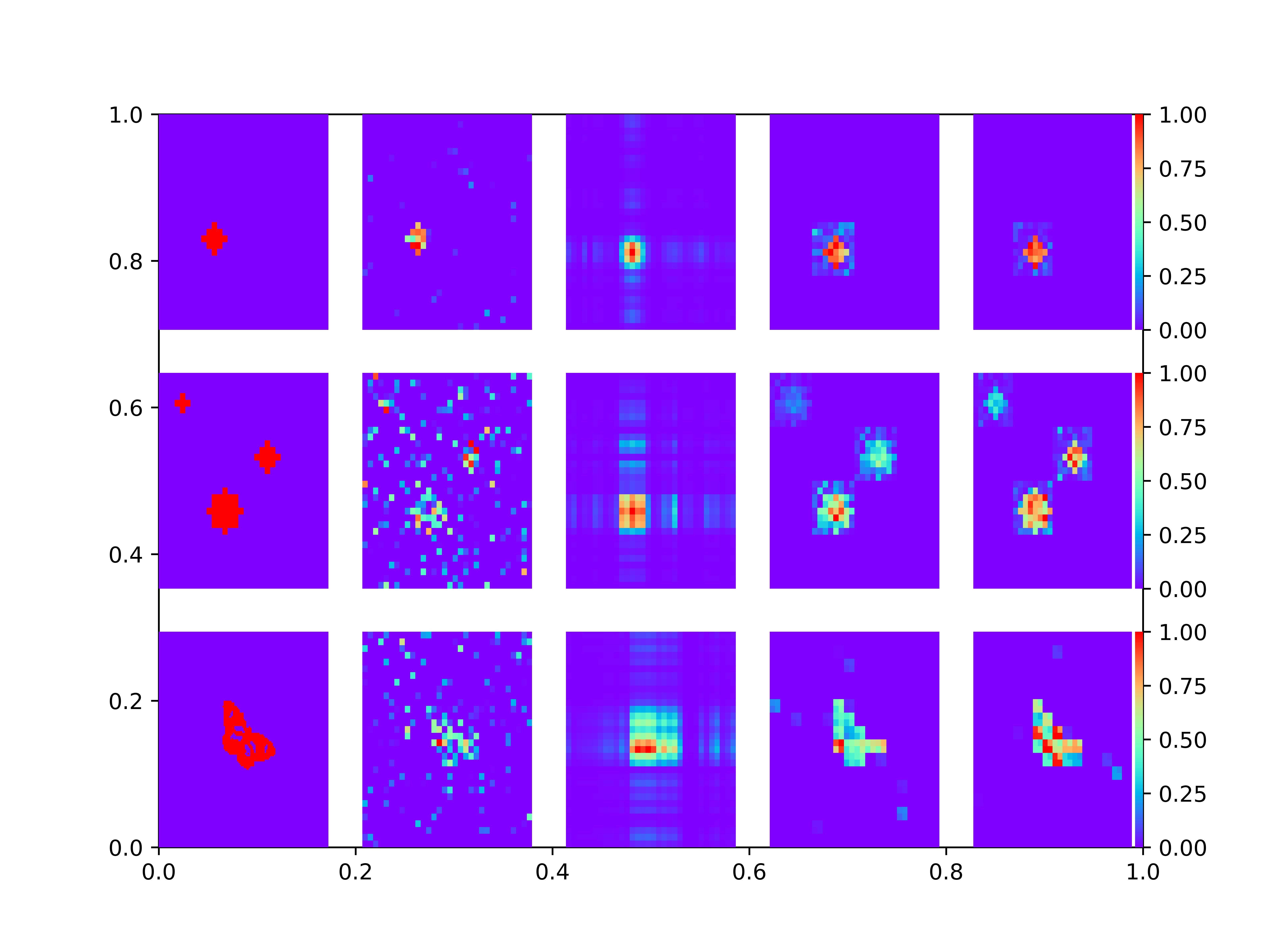}
	\caption{An illustration of the estimated image coefficient $\hbC \in \mathbb{R}^{32 \times 32}$ in the simulation study with identity covariance  and $(\rho_1,\rho_2)=(0.7,0.5)$. From left to right columns: True signals; naive; TGCCA; 1-term; R-term.}
	\label{fig:boat1}
\end{figure}

Furthermore, we consider three different image coefficients $\bC$: ``1-block'', ``3-block'', and ``butterfly". The three coefficients representing ``single'', ``multiple'' and ``complex'' signal shapes, respectively. 
We let $\bC_{jk}=1$ if the pixel $(j,k)$ fall in the signal shape and $\bC_{jk}=0$. See Figure \ref{fig:boat1} for  illustrations of the signal shapes.  Regarding the genetic covariates $\btheta$, we fix its sparsity to be 5, i.e., $\|\btheta\|_0=5$ and randomly select 5 nonzero covariates. The non-zero covariate values are set to be equal, and $\btheta$ is normalized such that $\|\btheta\|_2=1$.


We implement the proposed approach with 1-term SKPD and R-term SKPD with $R$ tuned by BIC discussed earlier. The block size is fixed to be $(d_1,d_2)=(8,8)$, while the ``location indicator'' matrices $\bA_r$ are of size $(p_1,p_2)=(4,4)$. Under such a scenario, the ``1-block'' and ``3-block'' coefficients could be written into a Kronecker product form with $R=1$ and $R=3$ respectively. However, the ``butterfly'' coefficients could not be written into a Kronecker product form with limited ranks (e.g., $R\le 10$). Therefore, the ``butterfly'' signal also represents a mis-specified setting and it allows us to test the robustness of our approach.

We evaluate the region detection, feature selection, coefficients estimation performance along with the computation time of our approach. Specifically, we measure the (pixel-wise) True Positive Rate (TPR) and False Positive Rate (FPR) for the image coefficients, where
 \bes
 \text{FPR} =\sum_{i=1}^{D_1} \sum_{j=1}^{D_2} \frac{\boldsymbol{I}\left(\hbC_{i j} \neq 0\right) \cdot \boldsymbol{I}\left(\bC_{i j}=0\right)}{\boldsymbol{I}\left(\bC_{i j}=0\right)}, \  \ \ \ \text{TPR} =\sum_{i=1}^{D_1} \sum_{j=1}^{D_2} \frac{I\left(\hbC_{i j} \neq 0\right) \cdot \boldsymbol{I}\left(\bC_{i j} \neq 0\right)}{\boldsymbol{I}\left(\bC_{i j} \neq 0\right)}.
 \ees
Similarly, we also use TRP and FPR to evaluate the feature selection performance of the genetic coefficients. Moreover, we calculate the Mean Squared Error (MSE) to evaluate the estimation performance of $\bC$ and $\btheta$, i.e.,  $\|\bC-\hbC\|_2^2$ and $\|\btheta-\hbtheta\|_2^2$.

\textcolor{black}{The performance of our approach is compared with two CCA methods. The first is the naive sparse CCA (referred to naive), which could be viewed as a special SKPD with $d_1=d_2=1$. The second competing method is the Tensor Generalized Canonical Correlation Analysis (\citealt{girka2024tensor}, referred to TGCCA), where CPD is employed on the canonical coefficients to reduce tensor dimension. 
However, as TGCCA does not incorporate a sparsity constraint, the region detection and feature selection measures (such as TPR, FPR) could not be compared.
Consequently, we limit the comparisons to the estimation performance, i.e., MSE of the canonical coefficients.
\textcolor{black}{We set the rank value in TGCCA to one, which is the default value in the source code. It is worth noting that we also experimented with higher rank values and obtained similar results.} The simulation is repeated for 100 times and the mean results are reported in Table  \ref{tab3} for identity covariance, Table \ref{tab4} for Toeplitz covariance and Table \ref{tab7} for computation time.
In addition,  we also plot an estimated $\hbC$ using different methods in one repetition in Figure \ref{fig:boat1}.}

\begin{table}[h!]
    \centering
    \begin{tabular}{cccccccccc}
         \multicolumn{2}{c}{\text {Identity Covariance }} & \multicolumn{3}{c}{\text { TPR of $\boldsymbol{C}$ }} & \multicolumn{3}{c}{\text { FPR of $\boldsymbol{C}$ }}\\
\hline \text { Shape } & ($\rho_1$,$\rho_2$)  & \text { 1-term} & \text { R-term } & \text{Naive} &\text { 1-term } & \text { R-term} & \text{Naive}   \\
\multirow{3}{*} {1-block} & (0.8,0.8) &  \textbf{1.0} & \textbf{1.0} & {0.99}  & \textbf{0.05} & \textbf{0.05} & {0.12}  \\
& (0.8,0.6)  & \textbf{1.0} & \textbf{1.0}  & \textbf{1.0} & \textbf{0.05} & \textbf{0.05} & {0.14}  \\
& (0.7,0.5)  & \textbf{1.0} & \textbf{1.0} & \textbf{1.0} & \textbf{0.05} & \textbf{0.05}& {0.15} \\
\hline
\multirow{3}{*}{ 3-block } & (0.8,0.8)  & {0.92} & \textbf{0.96} & {0.88} &\textbf{0.12} &{0.15}& {0.15}  \\
& (0.8,0.6)  & {0.92} & \textbf{0.98}  & {0.96} &0.16 & \textbf{0.14} & {0.21} \\
& (0.7,0.5)  & {0.96}& \textbf{0.98}  & {0.88} &\textbf{0.15} & \textbf{0.15}& {0.22} \\
\hline
 \multirow{3}{*}{\text { Butterfly }} & (0.8,0.8)  & {0.85} & \textbf{0.94}  & {0.84} & \textbf{0.01} & {0.05} &  {0.20} \\
& (0.8,0.6)  & 0.96 & \textbf{0.97}  & {0.89} & {0.08} & \textbf{0.05} & {0.20} \\
& (0.7,0.5)  & {0.95} & \textbf{0.96}  & {0.79} & {0.09} & \textbf{0.08}& {0.22}   \\
\\
& & \multicolumn{3}{c}{\text { TPR of $\boldsymbol{\theta}$}}&\multicolumn{3}{c}{\text { FPR of $\boldsymbol{\theta}$}}\\
\hline
 \text { Shape } & ($\rho_1,\rho_2$)  & \text { 1-term} & \text{R-term} &\text{Naive} & \text { 1-term} & \text{R-term} & \text{Naive} \\
 \multirow{3}{*}{\text { 1-block }} & (0.8,0.8)  & {0.56} & {0.75} & \textbf{0.84} & \textbf{0.02} & {0.05} & {0.28} \\
& (0.8,0.6)  & \textbf{0.99} & \textbf{0.99} & {0.98} &{0.06} & \textbf{0.05} & {0.10}\\
& (0.7,0.5) & {0.99} & \textbf{1.0} & {0.95} &{0.09} & \textbf{0.07} & {0.14} \\
\hline
\multirow{3}{*}{3-block} & (0.8,0.8)& \textbf{0.43} & {0.40}&{0.21} &{0.38} &{0.14}&  \textbf{0.07}\\
& (0.8,0.6) &{0.78} & {0.75} & \textbf{0.80} &\textbf{0.13} & {0.15} & {0.21} \\
& (0.7,0.5) & \textbf{0.90} & {0.82} & {0.66}&\textbf{0.13} & {0.18}& {0.27} \\
\hline
\multirow{3}{*}{\text { Butterfly }} & (0.8,0.8) & {0.24} & \textbf{0.60}  & {0.43} & \textbf{0.05} & {0.17} & {0.16} \\
& (0.8,0.6)& {0.96} & \textbf{0.97}  & {0.82}& \textbf{0.06} & \textbf{0.06} & {0.20} \\
& (0.7,0.5)&\textbf{0.98} & {0.97} & {0.74} &\textbf{0.08} & {0.09} & {0.26} \\
\\
    \end{tabular}
	\begin{tabular}{cccccccccc}
		 \multicolumn{2}{c}{\text {Identity Covariance }} & \multicolumn{4}{c}{\text { MSE of $\boldsymbol{C}$ }} & \multicolumn{4}{c}{\text { MSE of $\boldsymbol{\theta}$ }}\\
		\hline 
		\text { Shape } & ($\rho_1,\rho_2$)  & \text { 1-term} & \text{R-term}& \text{Naive}& \text{TGCCA} & \text { 1-term} & \text{R-term}& \text{Naive} & \text{TGCCA}\\
		\multirow{3}{*}{1-block} & (0.8,0.8)  & \textbf{0.052} & 0.056& {0.075} & 1.655 &{0.813} &\textbf{0.805} & 0.940 & {1.629}\\
		& (0.8,0.6)  & \textbf{0.042} & {0.038}& {0.055} & 1.794 & \textbf{0.160} & {0.168}& 0.235 & 1.777\\
		& (0.7,0.5) &  \textbf{0.066} &{0.068}&{0.134} & 1.517 &{0.139} & \textbf{0.172} & 0.306 & 1.488\\
        \hline
		\multirow{3}{*}{3-block} & (0.8,0.8)& 0.487 & \textbf{0.412}&  0.496 & 1.944 & 1.864 &\textbf{1.691} & 1.894& {1.926}\\
		& (0.8,0.6) &0.347 & \textbf{0.329}& 0.353 & 1.608&{0.728}&\textbf{0.691} & 0.827 & 1.474 \\
		& (0.7,0.5) & 0.389 & \textbf{0.342}& 0.632 & 1.666 & 0.694&\textbf{0.551} & 1.063 & 1.559\\
        \hline
		\multirow{3}{*}{\text { Butterfly }} & (0.8,0.8) & {0.297} & \textbf{0.192}& 
		0.526  & 1.930 & {1.547} & \textbf{1.215} & 1.772 & 1.926 \\
		& (0.8,0.6)& 0.149 & \textbf{0.136}& 0.377 & 1.532 &0.267 & \textbf{0.252}& 0.793 & 1.482  \\
		& (0.7,0.5)&{0.173} & \textbf{0.162} & 0.681 & 1.335 &\textbf{0.224} & {0.262} & 1.028 & 1.276 \\
		
	\end{tabular}
    \vspace{0.1in}
	\caption{Region detection, feature selection and estimation performance of different methods under  identity covariance}
	\label{tab3}
\end{table}

\begin{table}[h!]
	\centering
	\begin{tabular}{cccccccc}
		 \multicolumn{2}{c}{\text {Toeplitz Covariance }} & \multicolumn{3}{c}{\text { TPR of $\boldsymbol{C}$ }} & \multicolumn{3}{c}{\text { FPR of $\boldsymbol{C}$ }}\\
		\hline \text { Shape } & ($\rho_1,\rho_2$)  & \text { 1-term} & \text { R-term } & \text{Naive} & \text { 1-term } & \text { R-term} & \text{Naive}  \\
		\multirow{3}{*}{1-block} & (0.8,0.8) &  \textbf{1.0} & \textbf{1.0} & {0.96} & \textbf{0.05}& \textbf{0.05} & {0.25} \\
		& (0.8,0.6)  & {0.95} & \textbf{1.0} &{0.92} & \textbf{0.05} & \textbf{0.05} & {0.26} \\
		& (0.7,0.5)  & {0.98} & \textbf{1.0} & {0.52}& \textbf{0.05} & 0.06& {0.42}  \\
        \hline
		\multirow{3}{*}{3-block} & (0.8,0.8)  & \textbf{0.96} & \textbf{0.96} &{0.72}  &0.20 & \textbf{0.12}& {0.25} \\
		& (0.8,0.6)  &{0.92} & \textbf{0.97} & {0.71} &\textbf{0.15} & \textbf{0.15}& {0.27} \\
		& (0.7,0.5)  & {0.94} & \textbf{0.95} & {0.63} &{0.18} & \textbf{0.13}& {0.36}  \\
		\hline
		\multirow{3}{*}{\text { Butterfly }} & (0.8,0.8)  & {0.88} & \textbf{0.90} & {0.72} & {0.14} & \textbf{0.03} & {0.15}  \\
		& (0.8,0.6)  & {0.91} & \textbf{0.94} &{0.60} & {0.15} & \textbf{0.04} & {0.20} \\
		& (0.7,0.5)  & {0.83} & \textbf{0.92} & {0.55} & {0.19}& \textbf{0.05}& {0.26} \\
		\\
        
		& & \multicolumn{3}{c}{\text { TPR of $\boldsymbol{\theta}$}}&\multicolumn{3}{c}{\text { FPR of $\boldsymbol{\theta}$}}\\
		\hline
		\text { Shape } & ($\rho_1,\rho_2$)  & \text { 1-term} & \text{R-term} &\text{Naive}& \text { 1-term} & \text{R-term} & \text{Naive}\\
		\multirow{3}{*}{1-block} & (0.8,0.8)  &\textbf{0.30} & \textbf{0.30}& {0.25} & {0.11} & {0.10} & \textbf{0.09}\\
		& (0.8,0.6)  & \textbf{0.75} &{0.71} & {0.49} & {0.16} & \textbf{0.14} &\textbf{0.14}\\
		& (0.7,0.5) & \textbf{0.73} & {0.66}& {0.38} & {0.17} & \textbf{0.16} & {0.20}\\
        \hline
		\multirow{3}{*}{3-block} & (0.8,0.8)&{0.20} & \textbf{0.25}& {0.16} &{0.11} &{0.07}& \textbf{0.06} \\
		& (0.8,0.6) &{0.56} & \textbf{0.57}& {0.40} &\textbf{0.10} & \textbf{0.10} & \textbf{0.10}  \\
		& (0.7,0.5) & {0.53} & \textbf{0.55}& {0.43} & {0.13} &{0.12}& \textbf{0.09} \\
       
		\multirow{3}{*}{\text { Butterfly }} & (0.8,0.8) & {0.27} & \textbf{0.28}& {0.19} & {0.06} & {0.06} & \textbf{0.05}  \\
		& (0.8,0.6)& \textbf{0.66} & {0.61} & {0.47} & {0.16} & {0.11} & \textbf{0.10}  \\
		& (0.7,0.5)&\textbf{0.74} & {0.67}& {0.49} &{0.19} &{0.13} & \textbf{0.10}\\
		\\
	\end{tabular}
	\begin{tabular}{cccccccccc}
		 \multicolumn{2}{c}{\text {Toeplitz Covariance }} & \multicolumn{4}{c}{\text { MSE of $\boldsymbol{C}$ }} & \multicolumn{4}{c}{\text { MSE of $\boldsymbol{\theta}$ }}\\
		\hline 
		\text { Shape } & ($\rho_1,\rho_2$)  & \text { 1-term} & \text{R-term}& \text{Naive}& \text{TGCCA} & \text { 1-term} & \text{R-term}& \text{Naive} & \text{TGCCA}\\
		\multirow{3}{*}{1-block} & (0.8,0.8)  & \textbf{0.656} & 0.669& {0.928} & 1.872 &{1.409} &\textbf{1.354} & 1.498 & 1.903 \\
		& (0.8,0.6)  & \textbf{0.618} & {0.650}& {0.906} & 1.806  & \textbf{1.040} & {0.819}& 1.226 & 1.847 \\
		& (0.7,0.5) & \textbf{1.003} & {1.030}&{1.379} & 1.642 &{1.008} & \textbf{0.986} & 1.255 & 1.745 \\
        \hline
		\multirow{3}{*}{3-block} & (0.8,0.8)& 1.186 & {1.104}&  \textbf{0.985} & 1.673  & {1.671} &\textbf{1.362} &1.558& 1.741\\
		& (0.8,0.6) &\textbf{0.963} & {1.020}& 1.115 & 1.930 &1.038&\textbf{0.933} & 1.119& 1.947  \\
		& (0.7,0.5) & 1.211 & \textbf{1.049}& 1.748 & 1.707 & 1.062&\textbf{1.048} & 1.116 & 1.760\\
        \hline
		\multirow{3}{*}{\text { Butterfly }} & (0.8,0.8) & \textbf{0.512} & 0.585& 
		0.664  & 1.561 & {1.384} & \textbf{1.353} & 1.386 & 1.740 \\
		& (0.8,0.6)& \textbf{0.507} & {0.617}& 0.837 &  1.510&\textbf{0.922} & {0.950}& 1.101 & 1.714  \\
		& (0.7,0.5)&{0.798} &\textbf{0.625} & 1.198 &  1.845&{1.008} & \textbf{0.760} & 1.028 & 1.913  \\
		
	\end{tabular}
    \vspace{0.1in}
	\caption{Region detection, feature selection and estimation performance of different methods under  Toeplitz covariance}
	\label{tab4}
\end{table}

\begin{table}[h!]
	\centering
	\begin{tabular}{cccc}
		{\text {1-term }} & {\text {R-term}} & {\text { Naive}} & {\text{TGCCA}}\\
		\hline 1.9 & 6.0 & 23.0 & 2.4    \\

	\end{tabular}
	\vspace{0.1in}
	\caption{Computation times of different methods (in seconds)}
	\label{tab7}
\end{table}

By  Table \ref{tab3}, Table \ref{tab4}, the proposed approach with both 1-term and R-term SKPD demonstrate competitive performance on all the metrics. Under most combinations of signal shapes, correlations $(\rho_1,\rho_2)$, and covariance matrices, our method is capable of recovering more than 95\% of the true image signals and genetic features while simultaneously maintaining a low false positive rate. As a comparison, the naive approach that treat pixels independently often exhibits inferior performance in not only image region detection, but also genetics feature selection. This demonstrate the advantages of ``block-wise'' detection as in SKPD comparing to the ``pixel-wise'' selection. When comparing the MSE with TGCCA, our approach also demonstrate significant advantages.
In particular,  when the true signal is ``three-circle", TGCCA fails to recover any of the three circles. As a comparison, both 1-term and R-term SKPD could accurately locate all three signals. One of the major reasons is that we are considering sparse image and genetic coefficients, which could not be effectively addressed by TGCCA.
If we further compare the 1-term and R-term SKPDs, Table \ref{tab3} and Table \ref{tab4} suggest that with a 1-block signal shape, the performances of the 1-term and R-term methods are similar. However, when dealing with a ``3-block'' or a ``butterfly'' signal, the R-term SKPD generally outperforms the 1-term method.

The simulation study was conducted on a standard personal computer equipped with an Intel Core i5 CPU. Table \ref{tab7} summarizes the computation times required. It suggests that our approach significantly reduces the computation time over the naive sparse CCA, and it is also comparable to the TGCCA. By introducing the Kronecker product structure, we can not only capture blockwise signals, but also significantly reduce computational complexities.

\section{Conclusion}

We propose a general framework to joint modeling brain imaging, genetics information and clinical phenotype outcomes. To the best of our knowledge, this is the first framework that could address real high-resolution images, detect significant brain regions and genetic variants.
The proposed approach is  based on a multi-block canonical correlation analysis and a sparse Kronecker decomposition. Such a decomposition could significantly reduce image dimensions, substantially alleviate the computational burdens and address this ``big data squared'' problem. Consequently,  real MRI could be directly analyzed without relying on derived quantitative traits or summarized statistics. The efficacy of our method is well demonstrated through extensive simulation studies.


In our analysis of the UK Biobank dataset, which includes brain MRI images, genetic data, and clinical phenotypes, we have identified significant associations between reaction time, the caudate nucleus brain region, and several biomarkers. These findings, which are consistent across multiple independent data batches, not only align with existing literature but also provide valuable insights into the study of the pathway from biomarkers to brain function and ultimately to cognitive behaviors.

This paper is limited to CCA with linear functions of the genetics and imaging input. It is of our future interest to generalize the linear requirement to nonlinear functions through neural networks. Over the past decade, neural networks have demonstrated their immense capabilities in a wide range of applications, including image analysis and language modeling. However, a significant concern with neural network is that they lack of necessary interpretability. In particular, in our situation, it is crucial to understand which brain regions are linked to the phenotype and their associations with particular (sets of) SNPs. Consequently, a key challenge lies in expanding this framework to incorporate nonlinear models while simultaneously achieve variable selection and region detection. In the literature,  many efforts have been devoted to improve the interpretability of neural networks through different perspectives. An particular interesting work is the LassoNet \citep{lemhadri2021lassonet}. By adding a skip (residual) layer to constrain the participation of any hidden layer, LassoNet achieves global feature selection. In the future, it is of our great interest to 
borrow the idea of LassoNet for nonlinear modeling and broadening its application to the realm of imaging genetics.

\bibliographystyle{apalike}  
\bibliography{name} 

\begin{thebibliography}{}

\bibitem[Brouwer et~al., 2022]{brouwer2022}
Brouwer, R.~M., Klein, M., Grasby, K.~L., Schnack, H.~G., Jahanshad, N., Teeuw, J., Thomopoulos, S.~I., Sprooten, E., Franz, C.~E., Gogtay, N., et~al. (2022).
\newblock Genetic variants associated with longitudinal changes in brain structure across the lifespan.
\newblock {\em Nature neuroscience}, 25(4):421--432.

\bibitem[Bycroft et~al., 2018]{Bycroft2018}
Bycroft, C., Freeman, C., and Petkova, D. e.~a. (2018).
\newblock The uk biobank resource with deep phenotyping and genomic data.
\newblock {\em Nature}, 562:203--209.

\bibitem[Cai et~al., 2020]{cai2020minimal}
Cai, N., Revez, J.~A., Adams, M.~J., Andlauer, T.~F., Breen, G., Byrne, E.~M., Clarke, T.-K., Forstner, A.~J., Grabe, H.~J., Hamilton, S.~P., et~al. (2020).
\newblock Minimal phenotyping yields genome-wide association signals of low specificity for major depression.
\newblock {\em Nature genetics}, 52(4):437--447.

\bibitem[Chi et~al., 2013]{chi2013imaging}
Chi, E.~C., Allen, G.~I., Zhou, H., Kohannim, O., Lange, K., and Thompson, P.~M. (2013).
\newblock Imaging genetics via sparse canonical correlation analysis.
\newblock In {\em 2013 IEEE 10th International Symposium on Biomedical Imaging}, pages 740--743. IEEE.

\bibitem[Consortium et~al., 2011]{MS2011}
Consortium, I. M. S.~G., 2, W. T. C. C.~C., Sawcer, S., Hellenthal, G., Pirinen, M., Spencer, C.~C., Patsopoulos, N.~A., Moutsianas, L., Dilthey, A., Su, Z., Freeman, C., Hunt, S.~E., Edkins, S., Gray, E., Booth, D.~R., Potter, S.~C., Goris, A., Band, G., Oturai, A.~B., Strange, A., and …~Compston, A. (2011).
\newblock Genetic risk and a primary role for cell-mediated immune mechanisms in multiple sclerosis.
\newblock {\em Nature}, 476(7359):214--219.

\bibitem[Consortium*† et~al., 2019]{MS2019}
Consortium*†, I. M. S.~G., ANZgene, IIBDGC, and WTCCC2 (2019).
\newblock Multiple sclerosis genomic map implicates peripheral immune cells and microglia in susceptibility.
\newblock {\em Science}, 365(6460):eaav7188.

\bibitem[Davies et~al., 2018]{davies2018}
Davies, G., Lam, M., Harris, S.~E., Trampush, J.~W., Luciano, M., Hill, W.~D., Hagenaars, S.~P., Ritchie, S.~J., Marioni, R.~E., Fawns-Ritchie, C., et~al. (2018).
\newblock Study of 300,486 individuals identifies 148 independent genetic loci influencing general cognitive function.
\newblock {\em Nature communications}, 9(1):2098.

\bibitem[Davies et~al., 2016]{davies2016genome}
Davies, G., Marioni, R.~E., Liewald, D.~C., Hill, W.~D., Hagenaars, S.~P., Harris, S.~E., Ritchie, S.~J., Luciano, M., Fawns-Ritchie, C., Lyall, D., et~al. (2016).
\newblock Genome-wide association study of cognitive functions and educational attainment in uk biobank (n= 112 151).
\newblock {\em Molecular psychiatry}, 21(6):758--767.

\bibitem[Doornik and Hansen, 1994]{door}
Doornik, J. and Hansen, H. (1994).
\newblock Non-linear registration aka spatial normalisation. internal technical report tr07ja2.
\newblock {\em Oxford Centre for Functional Magnetic Resonance Imaging of the Brain,Department of Clinical Neurology, Oxford University}.

\bibitem[Driscoll et~al., 2020]{driscoll2020neuroanatomy}
Driscoll, M.~E., Bollu, P.~C., and Tadi, P. (2020).
\newblock Neuroanatomy, nucleus caudate.

\bibitem[Fan and Lv, 2008]{FanLv08}
Fan, J. and Lv, J. (2008).
\newblock Sure independence screening for ultrahigh dimensional feature space.
\newblock {\em Journal of the Royal Statistical Society: Series B (Statistical Methodology)}, 70(5):849--911.

\bibitem[Ge et~al., 2012]{ge2012increasing}
Ge, T., Feng, J., Hibar, D.~P., Thompson, P.~M., and Nichols, T.~E. (2012).
\newblock Increasing power for voxel-wise genome-wide association studies: the random field theory, least square kernel machines and fast permutation procedures.
\newblock {\em Neuroimage}, 63(2):858--873.

\bibitem[Giannakopoulou et~al., 2021]{giannakopoulou2021}
Giannakopoulou, O., Lin, K., Meng, X., Su, M.-H., Kuo, P.-H., Peterson, R.~E., Awasthi, S., Moscati, A., Coleman, J.~R., Bass, N., et~al. (2021).
\newblock The genetic architecture of depression in individuals of east asian ancestry: a genome-wide association study.
\newblock {\em JAMA psychiatry}, 78(11):1258--1269.

\bibitem[Girka et~al., 2024]{girka2024tensor}
Girka, F., Gloaguen, A., Le~Brusquet, L., Zujovic, V., and Tenenhaus, A. (2024).
\newblock Tensor generalized canonical correlation analysis.
\newblock {\em Information Fusion}, 102:102045.

\bibitem[Goldsmith et~al., 2014]{goldsmith2014smooth}
Goldsmith, J., Huang, L., and Crainiceanu, C.~M. (2014).
\newblock Smooth scalar-on-image regression via spatial bayesian variable selection.
\newblock {\em Journal of Computational and Graphical Statistics}, 23(1):46--64.

\bibitem[Grabner et~al., 2006]{grab}
Grabner, G., Janke, A.~L., Budge, M.~M., Smith, D., Pruessner, J., and Collins, D.~L. (2006).
\newblock Symmetric atlasing and model based segmentation: An application to the hippocampus in older adults.
\newblock In {\em Medical Image Computing and Computer-Assisted Intervention -- MICCAI 2006}, pages 58--66. Springer Berlin Heidelberg.

\bibitem[Grahn et~al., 2008a]{grahn2008cognitive}
Grahn, J.~A., Parkinson, J.~A., and Owen, A.~M. (2008a).
\newblock The cognitive functions of the caudate nucleus.
\newblock {\em Progress in neurobiology}, 86(3):141--155.

\bibitem[Grahn et~al., 2008b]{grahn2008}
Grahn, J.~A., Parkinson, J.~A., and Owen, A.~M. (2008b).
\newblock The cognitive functions of the caudate nucleus.
\newblock {\em Progress in neurobiology}, 86(3):141--155.

\bibitem[Grahn et~al., 2009]{grahn2009}
Grahn, J.~A., Parkinson, J.~A., and Owen, A.~M. (2009).
\newblock The role of the basal ganglia in learning and memory: neuropsychological studies.
\newblock {\em Behavioural brain research}, 199(1):53--60.

\bibitem[Hanafi and Kiers, 2006]{hanafi2006analysis}
Hanafi, M. and Kiers, H.~A. (2006).
\newblock Analysis of k sets of data, with differential emphasis on agreement between and within sets.
\newblock {\em Computational Statistics \& Data Analysis}, 51(3):1491--1508.

\bibitem[Hawi et~al., 2018]{hawi2018case}
Hawi, Z., Yates, H., Pinar, A., Arnatkeviciute, A., Johnson, B., Tong, J., Pugsley, K., Dark, C., Pauper, M., Klein, M., et~al. (2018).
\newblock A case--control genome-wide association study of adhd discovers a novel association with the tenascin r (tnr) gene.
\newblock {\em Translational psychiatry}, 8(1):284.

\bibitem[Huang et~al., 2015]{huang2015fvgwas}
Huang, M., Nichols, T., Huang, C., Yu, Y., Lu, Z., Knickmeyer, R.~C., Feng, Q., Zhu, H., Initiative, A. D.~N., et~al. (2015).
\newblock Fvgwas: Fast voxelwise genome wide association analysis of large-scale imaging genetic data.
\newblock {\em Neuroimage}, 118:613--627.

\bibitem[Jenkinson et~al., 2002]{jenkin}
Jenkinson, M., Bannister, P., Brady, M., and Smith, S. (2002).
\newblock Improved optimization for the robust and accurate linear registration and motion correction of brain images.
\newblock {\em NeuroImage}, 17(2):825--841.

\bibitem[Jiji et~al., 2013]{jiji2013}
Jiji, S., Smitha, K.~A., Gupta, A.~K., Pillai, V. P.~M., and Jayasree, R.~S. (2013).
\newblock Segmentation and volumetric analysis of the caudate nucleus in alzheimer's disease.
\newblock {\em European journal of radiology}, 82(9):1525--1530.

\bibitem[Kang et~al., 2018]{kang2018scalar}
Kang, J., Reich, B.~J., and Staicu, A.-M. (2018).
\newblock Scalar-on-image regression via the soft-thresholded gaussian process.
\newblock {\em Biometrika}, 105(1):165--184.

\bibitem[Kendall et~al., 2019]{kendall2019cognitive}
Kendall, K.~M., Bracher-Smith, M., Fitzpatrick, H., Lynham, A., Rees, E., Escott-Price, V., Owen, M.~J., O'Donovan, M.~C., Walters, J.~T., and Kirov, G. (2019).
\newblock Cognitive performance and functional outcomes of carriers of pathogenic copy number variants: analysis of the uk biobank.
\newblock {\em The British Journal of Psychiatry}, 214(5):297--304.

\bibitem[Kettenring, 1971]{kettenring1971canonical}
Kettenring, J.~R. (1971).
\newblock Canonical analysis of several sets of variables.
\newblock {\em Biometrika}, 58(3):433--451.

\bibitem[Knutson et~al., 2001a]{knutson2001anticipation}
Knutson, B., Adams, C.~M., Fong, G.~W., and Hommer, D. (2001a).
\newblock Anticipation of increasing monetary reward selectively recruits nucleus accumbens.
\newblock {\em The Journal of neuroscience}, 21(16):RC159.

\bibitem[Knutson et~al., 2001b]{knutson2001dissociation}
Knutson, B., Fong, G.~W., Adams, C.~M., Varner, J.~L., and Hommer, D. (2001b).
\newblock Dissociation of reward anticipation and outcome with event-related fmri.
\newblock {\em Neuroreport}, 12(17):3683--3687.

\bibitem[Lange et~al., 2021]{lange2021nonconvex}
Lange, K., Won, J.-H., Landeros, A., and Zhou, H. (2021).
\newblock Nonconvex optimization via mm algorithms: Convergence theory.
\newblock In {\em Computational statistics in data science, American Cancer Society,}, pages 1--22.

\bibitem[Latourelle et~al., 2009]{latourelle2009}
Latourelle, J.~C., Pankratz, N., Dumitriu, A., Wilk, J.~B., Goldwurm, S., Pezzoli, G., Mariani, C.~B., DeStefano, A.~L., Halter, C., Gusella, J.~F., et~al. (2009).
\newblock Genomewide association study for onset age in parkinson disease.
\newblock {\em BMC medical genetics}, 10:1--14.

\bibitem[Lemhadri et~al., 2021]{lemhadri2021lassonet}
Lemhadri, I., Ruan, F., Abraham, L., and Tibshirani, R. (2021).
\newblock Lassonet: A neural network with feature sparsity.
\newblock {\em The Journal of Machine Learning Research}, 22(1):5633--5661.

\bibitem[Li et~al., 2021]{li2021super}
Li, T., Hu, J., Wang, S., and Zhang, H. (2021).
\newblock Super-variants identification for brain connectivity.
\newblock {\em Human brain mapping}, 42(5):1304--1312.

\bibitem[Liu et~al., 2022]{Liu2022}
Liu, Z., Yang, N., Dong, J., Tian, W., Chang, L., Ma, J., Guo, J., Tan, J., Dong, A., He, K., Zhou, J., Cinar, R., Wu, J., Salinas, A.~G., Sun, L., Kumar, M., Sullivan, B.~T., Oldham, B.~B., Pitz, V., Makarious, M.~B., and …~Tang, B. (2022).
\newblock Deficiency in endocannabinoid synthase daglb contributes to early onset parkinsonism and murine nigral dopaminergic neuron dysfunction.
\newblock {\em Nature communications}, 13(1):3490.

\bibitem[Luan et~al., 2011]{Luan2011}
Luan, Z., Zhang, Y., Lu, T., Ruan, Y., Zhang, H., Yan, J., Li, L., Sun, W., Wang, L., Yue, W., and Zhang, D. (2011).
\newblock Positive association of the human ston2 gene with schizophrenia.
\newblock {\em Neuroreport}, 22(6):288--293.

\bibitem[Luciano et~al., 2018]{luciano2018}
Luciano, M., Hagenaars, S.~P., Davies, G., Hill, W.~D., Clarke, T.-K., Shirali, M., Harris, S.~E., Marioni, R.~E., Liewald, D.~C., Fawns-Ritchie, C., et~al. (2018).
\newblock Association analysis in over 329,000 individuals identifies 116 independent variants influencing neuroticism.
\newblock {\em Nature genetics}, 50(1):6--11.

\bibitem[Lyall et~al., 2017]{lyall2017associations}
Lyall, D.~M., Celis-Morales, C.~A., Anderson, J., Gill, J.~M., Mackay, D.~F., McIntosh, A.~M., Smith, D.~J., Deary, I.~J., Sattar, N., and Pell, J.~P. (2017).
\newblock Associations between single and multiple cardiometabolic diseases and cognitive abilities in 474 129 uk biobank participants.
\newblock {\em European heart journal}, 38(8):577--583.

\bibitem[Lyall et~al., 2022]{lyall2022quantifying}
Lyall, D.~M., Quinn, T., Lyall, L.~M., Ward, J., Anderson, J.~J., Smith, D.~J., Stewart, W., Strawbridge, R.~J., Bailey, M.~E., and Cullen, B. (2022).
\newblock Quantifying bias in psychological and physical health in the uk biobank imaging sub-sample.
\newblock {\em Brain communications}, 4(3):fcac119.

\bibitem[Min et~al., 2019]{min2019tensor}
Min, E.~J., Chi, E.~C., and Zhou, H. (2019).
\newblock Tensor canonical correlation analysis.
\newblock {\em Stat}, 8(1):e253.

\bibitem[Nathoo et~al., 2019]{nathoo2019review}
Nathoo, F.~S., Kong, L., Zhu, H., and Initiative, A. D.~N. (2019).
\newblock A review of statistical methods in imaging genetics.
\newblock {\em Canadian Journal of Statistics}, 47(1):108--131.

\bibitem[Oertel-Kn{\"o}chel et~al., 2015]{oertel2015schizophrenia}
Oertel-Kn{\"o}chel, V., Lancaster, T.~M., Kn{\"o}chel, C., St{\"a}blein, M., Storchak, H., Reinke, B., Jurcoane, A., Kniep, J., Prvulovic, D., Mantripragada, K., et~al. (2015).
\newblock Schizophrenia risk variants modulate white matter volume across the psychosis spectrum: evidence from two independent cohorts.
\newblock {\em NeuroImage: Clinical}, 7:764--770.

\bibitem[Rovira et~al., 2020]{rovira2020}
Rovira, P., Demontis, D., S{\'a}nchez-Mora, C., Zayats, T., Klein, M., Mota, N.~R., Weber, H., Garcia-Mart{\'\i}nez, I., Pagerols, M., Vilar-Rib{\'o}, L., et~al. (2020).
\newblock Shared genetic background between children and adults with attention deficit/hyperactivity disorder.
\newblock {\em Neuropsychopharmacology}, 45(10):1617--1626.

\bibitem[Rudin et~al., 1992]{rudin1992nonlinear}
Rudin, L.~I., Osher, S., and Fatemi, E. (1992).
\newblock Nonlinear total variation based noise removal algorithms.
\newblock {\em Physica D: Nonlinear Phenomena}, 60(1-4):259--268.

\bibitem[Schrimsher et~al., 2002]{schrimsher2002}
Schrimsher, G.~W., Billingsley, R.~L., Jackson, E.~F., and Moore, B.~D. (2002).
\newblock Caudate nucleus volume asymmetry predicts attention-deficit hyperactivity disorder (adhd) symptomatology in children.
\newblock {\em Journal of Child Neurology}, 17(12):877--884.

\bibitem[Smith, 2002]{smith}
Smith, S.~M. (2002).
\newblock Fast robust automated brain extraction.
\newblock {\em Human Brain Mapping}, 17(3):143--155.

\bibitem[Stein et~al., 2010]{stein2010voxelwise}
Stein, J.~L., Hua, X., Lee, S., Ho, A.~J., Leow, A.~D., Toga, A.~W., Saykin, A.~J., Shen, L., Foroud, T., Pankratz, N., et~al. (2010).
\newblock Voxelwise genome-wide association study (vgwas).
\newblock {\em neuroimage}, 53(3):1160--1174.

\bibitem[Takase et~al., 2004]{takase2004}
Takase, K., Tamagaki, C., Okugawa, G., Nobuhara, K., Minami, T., Sugimoto, T., Sawada, S., and Kinoshita, T. (2004).
\newblock Reduced white matter volume of the caudate nucleus in patients with schizophrenia.
\newblock {\em Neuropsychobiology}, 50(4):296--300.

\bibitem[Tibshirani et~al., 2005]{tibshirani2005sparsity}
Tibshirani, R., Saunders, M., Rosset, S., Zhu, J., and Knight, K. (2005).
\newblock Sparsity and smoothness via the fused lasso.
\newblock {\em Journal of the Royal Statistical Society: Series B (Statistical Methodology)}, 67(1):91--108.

\bibitem[Trubetskoy et~al., 2022]{trubetskoy2022}
Trubetskoy, V., Pardi{\~n}as, A.~F., Qi, T., Panagiotaropoulou, G., Awasthi, S., Bigdeli, T.~B., Bryois, J., Chen, C.-Y., Dennison, C.~A., Hall, L.~S., et~al. (2022).
\newblock Mapping genomic loci implicates genes and synaptic biology in schizophrenia.
\newblock {\em Nature}, 604(7906):502--508.

\bibitem[Tseng, 2001]{tseng2001convergence}
Tseng, P. (2001).
\newblock Convergence of a block coordinate descent method for nondifferentiable minimization.
\newblock {\em Journal of optimization theory and applications}, 109:475--494.

\bibitem[Van~de Geer, 1984]{van1984linear}
Van~de Geer, J.~P. (1984).
\newblock Linear relations among k sets of variables.
\newblock {\em Psychometrika}, 49:79--94.

\bibitem[Vounou et~al., 2010]{vounou2010discovering}
Vounou, M., Nichols, T.~E., Montana, G., Initiative, A. D.~N., et~al. (2010).
\newblock Discovering genetic associations with high-dimensional neuroimaging phenotypes: A sparse reduced-rank regression approach.
\newblock {\em Neuroimage}, 53(3):1147--1159.

\bibitem[Wang et~al., 2012]{wang2012identifying}
Wang, H., Nie, F., Huang, H., Kim, S., Nho, K., Risacher, S.~L., Saykin, A.~J., Shen, L., and Initiative, A. D.~N. (2012).
\newblock Identifying quantitative trait loci via group-sparse multitask regression and feature selection: an imaging genetics study of the adni cohort.
\newblock {\em Bioinformatics}, 28(2):229--237.

\bibitem[Wang et~al., 2017]{wang2017generalized}
Wang, X., Zhu, H., and Initiative, A. D.~N. (2017).
\newblock Generalized scalar-on-image regression models via total variation.
\newblock {\em Journal of the American Statistical Association}, 112(519):1156--1168.

\bibitem[Wu and Feng, 2023]{wu2023sparse}
Wu, S. and Feng, L. (2023).
\newblock Sparse kronecker product decomposition: a general framework of signal region detection in image regression.
\newblock {\em Journal of the Royal Statistical Society Series B: Statistical Methodology}, 85(3):783--809.

\bibitem[Xu and Yin, 2013]{xu2013block}
Xu, Y. and Yin, W. (2013).
\newblock A block coordinate descent method for regularized multiconvex optimization with applications to nonnegative tensor factorization and completion.
\newblock {\em SIAM Journal on imaging sciences}, 6(3):1758--1789.

\bibitem[Yu et~al., 2022]{yu2022mapping}
Yu, D., Wang, L., Kong, D., and Zhu, H. (2022).
\newblock Mapping the genetic-imaging-clinical pathway with applications to alzheimer’s disease.
\newblock {\em Journal of the American Statistical Association}, 117(540):1656--1668.

\bibitem[Zhao et~al., 2023a]{zhao2023heart}
Zhao, B., Li, T., Fan, Z., Yang, Y., Shu, J., Yang, X., Wang, X., Luo, T., Tang, J., Xiong, D., et~al. (2023a).
\newblock Heart-brain connections: Phenotypic and genetic insights from magnetic resonance images.
\newblock {\em Science}, 380(6648):abn6598.

\bibitem[Zhao et~al., 2023b]{Zhu2023science}
Zhao, B., Li, T., Fan, Z., Yang, Y., Shu, J., Yang, X., Wang, X., Luo, T., Tang, J., Xiong, D., Wu, Z., Li, B., Chen, J., Shan, Y., Tomlinson, C., Zhu, Z., Li, Y., Stein, J.~L., and Zhu, H. (2023b).
\newblock Heart-brain connections: Phenotypic and genetic insights from magnetic resonance images.
\newblock {\em Science}, 380(6648).

\bibitem[Zhao et~al., 2021]{Zhu2021science}
Zhao, B., Li, T., Yang, Y., Wang, X., Luo, T., Shan, Y., Zhu, Z., Xiong, D., Hauberg, M.~E., Bendl, J., Fullard, J.~F., Roussos, P., Li, Y., Stein, J.~L., and Zhu, H. (2021).
\newblock Common genetic variation influencing human white matter microstructure.
\newblock {\em Science}, 372(6548).

\bibitem[Zhou et~al., 2013]{zhou2013tensor}
Zhou, H., Li, L., and Zhu, H. (2013).
\newblock Tensor regression with applications in neuroimaging data analysis.
\newblock {\em Journal of the American Statistical Association}, 108(502):540--552.

\bibitem[Zhu et~al., 2023]{zhu2023statistical}
Zhu, H., Li, T., and Zhao, B. (2023).
\newblock Statistical learning methods for neuroimaging data analysis with applications.
\newblock {\em Annual Review of Biomedical Data Science}, 6:73--104.

\end{thebibliography}
\newpage
\appendix
\section{The effect of initialization}
We have considered two random initialization procedures to investigate the robustness of our algorithm. The initial values of $\boldsymbol{\alpha}_r$ and $\boldsymbol{\beta}_r$ are randomly generated from a uniform distribution between (0.5,1), or from a normal distribution $N(1,0.1)$. We have compared these random initialization to the constant initialization under the simulation setting with a one-circle true signal shape, and the true canonical coefficients $(\rho_1,\rho_2) = (0.8,0.6)$. The region detection, feature selection and estimation results are summarized in the following Table 1. 
\begin{table}[h!]
    \centering
    \begin{tabular}{ccccccc}
          {\text {Initialization}} & {\text { TPR(C)}} & {\text { FPR(C)}} & {\text { TPR($\theta$)}} & {\text { FPR($\theta$)}} & {\text { MSE(C)}} & {\text { MSE($\theta$)}}\\
\hline 
Constant & 1.0 &  0.05 & 0.99 & 0.03 & 0.042 & 0.174    \\
Uniform & 1.0  & 0.05 & 0.98 & 0.03 & 0.041 & 0.156    \\
Normal & 1.0  & 0.05 & 1.0 & 0.03 & 0.040 & 0.146   \\

    \end{tabular}
    \caption{Region detection, feature selection and estimation performance of 1-term SKPD under different initialization.}
    \label{tab3}
\end{table}

By Table \ref{tab3}, it is clear that our algorithm performs robust towards different random initialization settings.
\\

\section{Data generation process}
We consider $N=1000$ subjects and each with an image $\boldsymbol{X}_i$ of size $32\times 32$, genetic covariates $\boldsymbol{z}_i$ of dimension $100$, and a scalar phenotype outcome $y_i$. Denote the population covariance matrix of vectorized image $ \boldsymbol{X}_i$ as $\boldsymbol{\Sigma}_x\in\mathbb{R}^{32^2\times 32^2}$, and the population covariance matrix of genetic covariates $\boldsymbol{z}_i$ as $\boldsymbol{\Sigma}_z\in\mathbb{R}^{100\times 100}$.

We consider two cases of the covariance structure of $\boldsymbol{\Sigma}_x$ and $\boldsymbol{\Sigma}_z$: identity covariance matrix and Toeplitz covariance matrix. For the identity case, we let $(\boldsymbol{\Sigma}_x,\boldsymbol{\Sigma}_z)=(\boldsymbol{I}_{32^2},\boldsymbol{I}_{100})$. While for the Toeplitz case, we let
$\boldsymbol{\Sigma}_x(j,k)=0.9^{|j-k|}$ for $j,k=1,\ldots, 32^2$ and $\boldsymbol{\Sigma}_z(j',k')=0.9^{|j'-k'|}$ for $ j',k'=1,\ldots, 100$. 

Denote the true canonical coefficient for image data $\boldsymbol{X}_i$ as $\boldsymbol{C}$, and true canonical coefficient for genetic covariates $\boldsymbol{z}_i$ as $\boldsymbol{\theta}$ respectively. To ensure the correlation between $\langle\boldsymbol{X},\boldsymbol{C}\rangle$ and $\langle \boldsymbol{z}, \boldsymbol{\theta} \rangle$ equals to a designed canonical correlation coefficient $\rho_1$, we generate $\boldsymbol{X}_i$ and $\boldsymbol{z}_i$ jointly from the following multivariate normal distribution:
$$\left(\begin{array}{l}\text{vec}({\boldsymbol{X}_i}) \\ \quad \boldsymbol{z}_i\end{array}\right) \sim \mathrm{N}\left(\left(\begin{array}{l}\boldsymbol{0} \\ \boldsymbol{0}\end{array}\right),\left(\begin{array}{cc}\boldsymbol{\Sigma}_x & \boldsymbol{\Sigma}_{xz} \\ \boldsymbol{\Sigma}_{xz}^T & \boldsymbol{\Sigma}_{z}\end{array}\right)\right)$$
where the cross-covariance matrix is modeled as $\boldsymbol{\Sigma}_{xz}=\rho_1\boldsymbol{\Sigma}_{x}\boldsymbol{\theta} \text{vec}(\boldsymbol{C})^{T}\boldsymbol{\Sigma}_{z}$.

In terms of the scalar phenotype variable $y_i$, its canonical coefficient is fixed as 1. Thus, we can generate $y_i$ directly without considering the probabilistic cross-covariance structure. Specifically, we aim to generate $y_i$ in a way that fixes the correlation between $y_i$ and $\langle\boldsymbol{X}_i, \boldsymbol{C}\rangle$ to a pre-determined canonical coefficient value of $\rho_2$. To achieve this objective, the following procedure is adopted to generate data vector $\boldsymbol{y} \in \mathbb{R}^{1000}$:
\begin{itemize}
    \item [1] Generate a random vector $\boldsymbol{y}^{*}$ from standard multivariate normal distribution with length $N$
    \item [2] Calculate vector $\boldsymbol{x}^{*}$ with length $N$ where the $i-th$ element is $\boldsymbol{x}^{*}_i$ = $\langle\boldsymbol{X}_i, \boldsymbol{C}\rangle$.  Implement linear regression $\boldsymbol{y}^* \sim \boldsymbol{x}^{*}$, and get the residual vector $\boldsymbol{x}^{\perp}$. By the property of ordinary least square method, this residual is orthogonal to $\boldsymbol{x}^*$
    \item [3] Obtain vector $\boldsymbol{y}$ of length $N$ as: $\boldsymbol{y}$ = $\rho_2 \operatorname{SD}\left(\boldsymbol{x}^{\perp}\right) \boldsymbol{x}^*+\sqrt{1-\rho_2^2} \operatorname{SD}(\boldsymbol{x}^*) \boldsymbol{x}^{\perp}$
\end{itemize}

It can be easily verified that the correlation between generated $\boldsymbol{y}$ and $\langle \boldsymbol{X}, \boldsymbol{C}\rangle$ is equal to the designed value $\rho_2$.
Through above generation process, we obtain three data blocks $\boldsymbol{X}_i$, $\boldsymbol{z}_i$ and $y_i$ such that the true canonical correlation between $\langle\boldsymbol{z}, \boldsymbol{\theta}\rangle$ and $\langle\boldsymbol{X}, \boldsymbol{C}\rangle$ is $\rho_1$ and the true canonical correlation between $\boldsymbol{y}$ and $\langle\boldsymbol{X}, \boldsymbol{C}\rangle$ is $\rho_2$. Once $\rho_1$ and $\rho_2$ are fixed, the true canonical correlation between $\boldsymbol{y}$ and $\langle\boldsymbol{z}, \boldsymbol{\theta}\rangle$ is also fixed.

\section{Further details on the UK Biobank data analysis} 
The presented figure illustrates the detected brain regions across ten independent batches in the analysis of UK Biobank data. It is evident that the region of detection exhibits consistency across the majority of the batches.

\begin{figure}[h!]
  \hspace{-0.5cm}\includegraphics[scale=0.45]{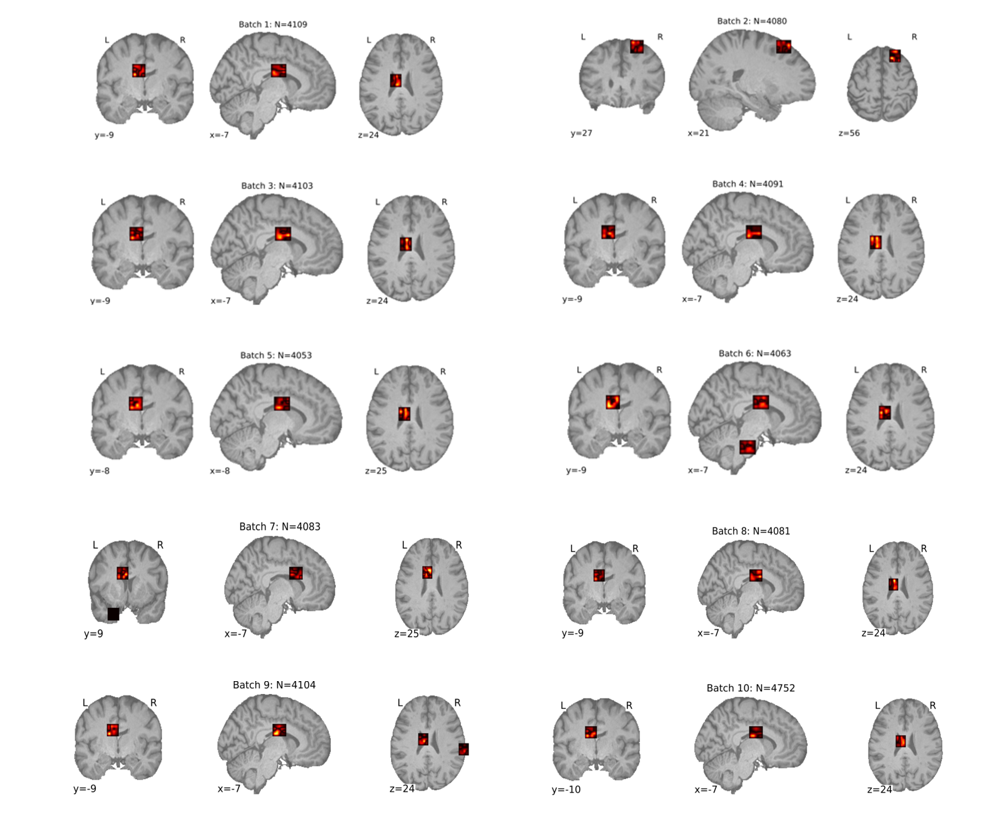}
  \caption{The presented figure illustrates the detected brain regions across ten independent batches in the analysis of UK Biobank data.}
\end{figure}

\end{document}